\documentstyle[12pt,graphicx,color,subfigure,epsfig,cite,amsmath,soul,ulem]{article}
\def\baselinestretch{1.3}
\newcommand{\ba}{\begin{array}}
\newcommand{\ea}{\end{array}}
\newcommand{\bd}{\begin{displaymath}}
\newcommand{\ed}{\end{displaymath}}
\newcommand{\be}{\begin{equation}}
\newcommand{\ee}{\end{equation}}
\newcommand{\bea}{\begin{eqnarray}}
\newcommand{\eea}{\end{eqnarray}}
\newcommand{\bei}{\begin{itemize}}
\newcommand{\eei}{\end{itemize}}

\def\a{\alpha}

\def\b{\beta}

\def\q2 {q^2}

\def\bt{\begin{table}}
\def\et{\end{table}}

\def\a{\alpha}

\def\b{\beta}

\def\q2 {q^2}

\def\bt{\begin{table}}
\def\et{\end{table}}

\catcode`@=11 
\def \gsim{\mathrel{\mathpalette\@versim>}}
\def \lsim{\mathrel{\mathpalette\@versim<}}
\def \@versim#1#2{\lower0.4ex\vbox{\baselineskip\z@skip\lineskip\z@skip
     \lineskiplimit\z@\ialign{$\m@th#1\hfil##\hfil$%
     \crcr#2\crcr\sim\crcr}}}
\catcode`@=12 

\def\issue(#1,#2,#3){{\bf #1}, #2 (#3)}

\def\epjc#1#2#3{{EPJC  \bf #1}, #3 (#2)}
\def\app#1#2#3{{Astropart. Phys.  \bf #1}, #3 (#2)}
\def\plb#1#2#3{{Phys. Lett. B  \bf #1}, #3 (#2)}
\def\jhep#1#2#3{{JHEP \bf #1}, #3 (#2)}

\def\prd#1#2#3{{Phys. Rev. D  \bf #1}, #3 (#2)}

\def\prl#1#2#3{{Phys. Rev. Lett.  \bf #1}, #3 (#2)}

\def\npb#1#2#3{{Nucl. Phys. B \bf #1}, #3 (#2)}

\def\PREP(#1,#2,#3){Phys.\ Rep. \issue(#1,#2,#3)}

\begin{document}

\begin{flushright}
OSU-HEP-13-6\\\
 UCRHEP-T535
\end{flushright}

\begin{center}
{\large \bf Non-universal SUGRA at LHC: Prospects and Discovery Potential}\\[15mm]
Subhaditya Bhattacharya\footnote{E-mail: subhaditya123@gmail.com}, Shreyashi Chakdar\footnote{E-mail: chakdar@okstate.edu}, 
Kirtiman Ghosh\footnote{E-mail: kirti.gh@gmail.com}, S. Nandi\footnote{E-mail: s.nandi@okstate.edu}\\[3 mm]
 1. { Department of Physics and Astronomy, \\
     University Of California, Riverside, CA 92501, USA}\\[2 mm]
2,3,4. { Department of Physics, Oklahoma State University, and
Oklahoma Center for High Energy Physics, Stillwater, OK 74078, USA}
\\[20mm]
\end{center}

\begin{abstract}
\noindent

We explore supersymmetry (SUSY) parameter space with non-universal high scale parameters in 
gravity mediated SUSY breaking (SUGRA) scenario that accommodates a Higgs mass of 
(125$\pm$2) GeV while satisfying cold dark matter relic density and other low energy constraints. 
We indicate a few benchmark points consistent with different dark matter annihilation processes 
where third family squarks are lighter than the first two as a requirement to keep the Higgs mass within 
the limit. We show that bottom rich and leptonic final states have better reach in such parameter space points 
and is the most likely scenario to discover SUSY at the upcoming run of LHC with center-of-mass energy 14 TeV.

\vskip 15pt
\noindent
\end{abstract}



\newpage

\setcounter{footnote}{0}

\def\baselinestretch{1.5}

\section{Introduction}

Supersymmetry (SUSY) \cite{SUSY,KaneKingRev} has been under scanner 
since last forty years or more. On-going Large Hadron Collider (LHC) has put strong 
bounds on the squark and gluino masses of minimal supersymmetric Standard Model (MSSM); 
particularly on minimal supergravity (mSUGRA) or constrained minimal 
supersymmetric Standard Model (CMSSM) \cite{msugra}, not seeing 
any of those supersymmetric particles. Still, SUSY search in different forms is the most 
studied subject of particle physics research due to its unparalleled theoretical appeal
and phenomenological implications.     

Out of different SUSY-breaking schemes, mSUGRA has been most popular 
due to its economy of parameters; the universal gaugino mass
($M_{1/2}$), the universal scalar mass ($m_0$), the universal
trilinear coupling ($A_0$) all at the GUT scale, $\tan\beta$, the
ratio of the vacuum expectation values (vev) of the two Higgses and
the sign of SUSY-conserving Higgsino mass parameter $\mu$. However,
this framework has been highly constrained by direct and indirect search experiments 
\cite{msgbound,b-jets-search,bsg-recent,mupmum} and 
non-universality in scalar \cite{3rd1,3rd2,3rd3,ucddOct2008,berez,Nath:1997qm,Cerdeno:2004zj,ellis-all,baer-all-non,so102,Datta:1999uh,BM-SB-AD2}
 and gaugino masses \cite{nonunigaugino,BM-AD-SB,sb-sn} are getting more and more importance 
 to keep low-scale SUSY alive.

Recent discovery of Higgs boson with $m_H\simeq$ 125 GeV at LHC 
by the ATLAS and CMS Collaborations \cite{higgs} has put a severe 
constraint on SUSY parameter space. SUSY Higgs gets significant correction 
from the top squark (stop) loop, which increases with increasing 
stop mixing and/or stop mass scale. Therefore, in order to get a Higgs boson 
around 125 GeV, significant stop mixing or a large stop mass scale is required. Large stop mixing 
results into large mass splitting in the stop sector and consequently gives rise to a lighter stop ($\tilde t_1$) 
in the mass spectrum. Hence, Higgs boson mass at 125 GeV results in a 
SUSY mass spectrum with light third family scalars.

 Light third family scalars, but relatively heavy first two families\footnote{Such scenarios 
 have already been considered for studies in different contexts \cite{3rd1,3rd2,3rd3}.} favor 
 SUSY discovery at future LHC runs given gluino ($\tilde g$) dominantly decays 
 into top-stop pairs ($\tilde g \to t \tilde t_1$) and subsequently stop decays into top-neutralino or b-chargino 
 where $t\rightarrow bW^\pm$ gives rise to multiple b-jets, leptons and large missing energy ($E_{T}\!\!\!\!/$). 
 Final states with multiple b jets and charged leptons, together with large missing energy,
 cut down the SM background much more than the usual SUSY signals with multijets plus large missing energy, and both 
 ATLAS and CMS experiments has achieved  b-tagging efficiency $50\%$ or more and have put bounds on 
 SUSY from the available data \cite{b-jets-search}. 
 
 Another important aspect of SUSY is the dark matter (DM); $R$-parity conservation yields a natural candidate namely, 
 the lightest supersymmetric particle (LSP). DM relic density limits from 
 WMAP \cite{WMAPdata} and PLANCK \cite{PLANCK} can be easily satisfied in the 
 non-universal gaugino and/or scalar mass scenarios where CMSSM is tightly constrained. 
 For example, if wino mass is smaller than bino mass at GUT scale ($M_2 \le M_1$), we obtain wino dominated LSP 
 yielding correct abundance in a larger parameter space. Similarly, non-universality in the scalar sector
  may results in a higgsino like LSP (from non-universality in the Higgs sector) or 
  stau-LSP co annihilation (from non-universality in the soft SUSY breaking stau mass). 
 We have systematically studied such non-universal gaugino  and/or scalar mass scenarios
 and proposed benchmark points for collider studies at LHC with $E_{CM}$= 14 TeV. 
 


Vast amount of work has already been done in mSUGRA to discover SUSY at the LHC. However, because of the 
observed Higgs mass, and the dark matter constraint, the only region left in mSUGRA and accessible at the LHC 
is the stop co-annihilation region (where the lighter top squark $\tilde{t_1}$ and the lightest neutralino $\tilde \chi_{1}^0$ annihilate to satisfy the
dark matter constraint. However, in this parameter space, $\tilde{t_1}$ mass is very close to the  $\tilde \chi_{1}^0$ mass
giving rise to very little high $p_T$ multijet activity from its decay \cite{degenerate-stop}. Significant number of works have also been done 
by increasing the number of parameters, with non-universal gaugino masses and non-unversality in the scalar masses satisfying all
the existing constraints \cite{recentworks}. However, we pin point that to survive Higgs mass and dark matter 
constraint in the framework of gravity mediated supersymmetry breaking, a larger region of parameter 
space is available with specific non-universal gaugino and scalar mass patterns with a generic signature in 
bottom rich, and bottom quark plus charged lepton rich final states with large missing energy, 
which with suitable cuts can be observed over the SM background at the 14 TeV LHC. 
We claim that these will be the most favorable final states at the 14 TeV LHC to discover SUSY or to put 
strongest bounds on them.

The paper is organized as follows. In Section 2, we discuss the model
under consideration and the selected benchmark points. We also review dark matter 
constraints on SUSY parameter space to motivate our benchmark points. In Section 3, we
discuss the  final states in which SUSY signals can be observed over the SM background, including the details of the
collider simulation strategy and the numerical results at the 14 TeV LHC. We conclude in Section 4.

\section {Model, Constraints and Benchmark Points}

\subsection{Constraints on SUSY models:} 
 

Following measurements play a key role to constrain SUSY parameter space. We discuss their effect 
and motivate how that leads eventually to the benchmark points chosen in this article for SUSY 
searches at LHC.

\begin{itemize}
\item
The main constraint on the SUSY parameter space after LHC 7/8 TeV data is that the 
CP even Higgs mass to be within  \cite{higgs}:

\begin{equation}
123 \le m_h \le 127.
\label{Higgs}
\end{equation}

\item
The branching ratio for $b \longrightarrow s \gamma$ \cite{bsg-recent} which
at the $3 \sigma$ level is
\begin{equation}
2.13 \times 10^{-4} < Br (b \rightarrow s \gamma) < 4.97 \times 10^{-4}.
\label{bsgammalimits}
\end{equation}

\item
We also take into account the constraint coming from $B_s  \longrightarrow \mu^+ \mu^-$ branching ratio 
which by LHCb observation \cite{mupmum} at 95\% CL is given as 
\begin{equation}
2 \times 10^{-9} < Br (B_s \rightarrow \mu^+ \mu^-) <  4.7 \times 10^{-9}.
\label{bstomumu}
\end{equation}

\item
Parameters are fine-tuned in a way that it gives a correct cold dark matter
relic abundance according to WMAP data \cite{WMAPdata},
which at $3 \sigma$ is
\begin{equation}
0.091 < \Omega_{CDM}h^2 < 0.128 \ ,
\label{relicdensity}
\end{equation}
where $\Omega_{CDM}$ is the dark matter
relic density in units of the critical
density and $h=0.71\pm0.026$ is the reduced Hubble constant
(namely, in units of
$100 \ \rm km \ \rm s^{-1}\ \rm Mpc^{-1}$).
\end{itemize}

To note here, the PLANCK 
constraints $0.112 \leq \Omega_{\rm DM} h^2 \leq 0.128$ \cite{PLANCK}
is more stringent, and cuts a significant amount of dark matter allowed 
SUSY parameter space. We choose our benchmark points satisfying 
PLANCK on top of WMAP.

In the following subsection, we discuss mainly the dark matter 
and Higgs mass constraints on SUSY parameter space as they have 
been the key to choose our benchmark points. 

\subsection{Dark matter and Higgs mass on SUSY: Benchmark Points}  

One of the main motivations for postulating $R$-parity conserving SUSY is the presence 
of a stable weakly interacting massive particle (WIMP) which can be a good cold dark matter.
Lightest neutralino $\tilde{\chi}_1^0$ is most often the lightest supersymmetric particle (LSP) 
and a good candidate for cold dark matter. In some regions of the parameter space, it has 
the annihilation cross-section to Standard Model (SM) particles yielding correct relic 
abundance satisfying WMAP/PLANCK \cite{WMAPdata,PLANCK}. 

In mSUGRA, $\tilde{\chi}_1^0$ is bino dominated in a large  
part of the parameter space. For a bino DM, WIMP miracle occurs when they annihilate to 
leptons via $t$-channel exchange of sleptons with mass in the 30-80 GeV range \cite{WIMP-miracle}. 
However, slepton masses that light was already discarded by direct slepton searches at LEP2 \cite{slepton-mass}. 
Therefore, after LEP2, some distinct parts of mSUGRA parameter space that satisfies relic abundance are as follows:
\begin{itemize}
\item {\bf The $h$-resonance region }\cite{h-resonance} is characterized by $2m_{\tilde{\chi_1^0}} \sim m_h$ which 
occurs at low $m_{1/2}$. In this region, $\tilde{\chi}_1^0$ annihilation cross-section enhances due to the presence of a $s$-channel $h$-resonance.
\item  {\bf $A$-funnel region}  \cite{A-funnel} is where $2m_{\tilde{\chi_1^0}} \sim m_A$; $A$ is the CP-odd Higgs boson. 
This region is characterized by large ${\rm tan}\beta \sim 50$.
\item {\bf Hyperbolic branch/focus point (HB/FP) region} \cite{HB/FP} is the parameter space where large 
$m_0$ region corresponds to small $\mu$ and thus Higgsino dominates $\tilde{\chi}_1^0$ and annihilates to $WW$, 
$ZZ$ and $Ah$ significantly. 
\item  {\bf Stau co-annihilation region}\cite{stau-co} arises if neutralino-LSP is nearly degenerate with the stau 
($m_{\tilde{\chi_1^0}} \simeq m_{\tilde \tau_1}$). In mSUGRA, this occurs at low $m_0$ and high $M_{1/2}$.
\item {\bf Stop co-annihilation} \cite{stop-co} occurs in mSUGRA with some particular values of $A_0$, 
where lighter stop ($\tilde t_1$) becomes nearly degenerate with the LSP.   
\end{itemize}

After LHC data with the discovery of Higgs and exclusion limits 
on the squark/gluino masses, many of the above DM regions in mSUGRA 
are highly constrained. With 20.3 fb$^{-1}$ integrated luminosity and $E_{CM}$= 8 TeV, 
ATLAS \cite{ATLAS} and CMS \cite{CMS} collaborations have excluded equal squark and 
gluino mass below 1.7 TeV completely ruling out  $h$-resonance region whereas, 
the $A$-funnel, stau and stop co-annihilation regions are partly excluded. 
Observation of Higgs mass at about 125 GeV indicates towards large $m_0$  ($m_0> 0.8$ TeV) 
and large $A_0$ ( $|A_0| > 1.8m_0$ for $m_0<5$ TeV) \cite{higgs-m0}. 
For $m_0>0.8$ TeV, stau co-annihilation is only viable at very large $M_{1/2}$ values 
which makes the SUSY discovery at the collider very challenging. 
The HB/FP region remains unscathed by the LHC squark/gluino searches 
as  it requires low $\mu$ at very large $m_0 \sim 3-10$ TeV for $A_0=0$. 
However, Higgs mass at 125 GeV (requires large $|A_0|$) push the 
region to much higher $m_0\sim 10-50$ TeV values. 
A small part of stop co-annihilation is the only region of mSUGRA 
parameter space alive, having some possibilities of seeing at 14 TeV LHC.

Non-universality in the gaugino and/or scalar sector on the other hand, 
can provide a lot more breathing space. The implications of direct search bound 
from LHC on neutralino dark matter have been studied extensively. See for example, \cite{DM-LHC1,DM-LHC2,AD-recent}. 
In our analysis, we choose four benchmark 
points (BP) which are motivated from different LSP annihilation and co-annihilation 
mechanism and consistent with all experimental limits. 

\begin{figure}[htbp]
\begin{center}
\centerline{\psfig{file=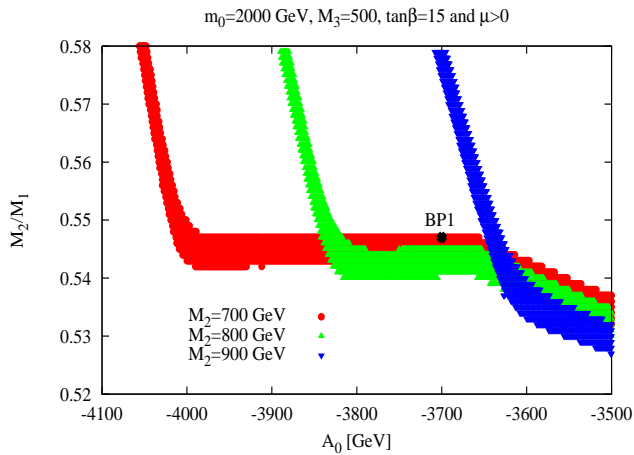,width=6.5 cm,height=8.5cm,angle=-90}}
\caption{A sample parameter space scan for gaugino mass non-universality with $M_3<M_2<M_1$ in $A_0$ vs $M_2/M_1$ 
plane to satisfy DM abundance. 
$M_2$= {(700, 800, 900)} GeV, yields three discrete consistent regions in red, blue and green respectively with $M_2/M_1$ varying along 
y-axis with $A_0$ varying along x-axis. We choose $M_3$= 500 GeV, $m_0$= 2000 GeV, $\tan\beta$= 15. BP1 represent a benchmark point of this sort.} 
\label{fig:BP1}
\end{center}
\end{figure}

\noindent {\bf i) BP1}:  If $M_2<M_1$ at GUT scale and EW scale, and $M_2< \mu$ at low scale, the LSP $\tilde{\chi}_1^0$ 
is wino dominated and then lightest chargino is almost degenerate with LSP. 
Chargino co-annihilation crucially controls relic abundance in such a region of parameter 
space, apart from larger wino component itself increases annihilation cross-section. A large part of purely 
wino DM hence provides under-abundance \cite{wino-DM}.  However, we scan the wino dominated 
parameter space where it is consistent with relic abundance from WMAP.
As an example, we have scanned the parameter space over $M_1,~M_2$ and $A_0$ for 
$m_0=2000$ GeV, $M_3=500$ GeV ${\rm tan}\beta=15$ and 
$\mu>0$. The allowed values of $M_2/M_1$ as a function of $A_0$ are plotted in 
Fig.~\ref{fig:BP1} for three different values $M_2=$ 700, 800 and 900 GeV in 
red, blue and green respectively. When we vary $M_2$ continuously, they merge 
into a continuous region. It is important to note in Fig.~\ref{fig:BP1} 
the vertical high $A_0$ region is dominated by stop co annihilation as the stop becomes 
lighter with increasing $A_0$ and a small change in $A_0$ results in a big change in $M_2/M_1$ to 
keep relic abundance within proper limit. The horizontal part of red, blue and green region 
with smaller $A_0$ on the other hand, represent wino dominated dark matter with nearly 
degenerate chargino and co-annihilation to yield proper abundance. 
For example, with $M_2=700$ GeV, $|A_0|>4000$ GeV is dominated by stop co-annihilation 
and $|A_0|<4000$ GeV characterizes wino DM. Our first benchmark point BP1 is a representative 
of this particular non-universal gaugino mass scenario $M_3<M_2<M_1$ with wino dominated DM. 
The benchmark points are explicitly written in Table \ref{tab:benchmark}. While 
gaugino mass non-universality has been used to obtain BP1, scalar masses are 
kept universal. 

Also note that gaugino non-universality with $M_3<M_2<M_1$ is obtained within the framework of
SUSY-GUT in $SU(5)$ or $SO(10)$ \cite{nonunigaugino, BM-AD-SB} with dimension five operator in the
extension of the gauge kinetic function $f_{\a\b}(\Phi^{j})$
\bea
 Re f_{\a \b}(\phi)F_{\mu \nu}^{\a}F^{\beta \mu \nu}= \frac {\eta (\Phi^s)}
{M}Tr(F_{\mu \nu}\Phi^N F^{\mu \nu})
\eea
\noindent
where non-singlet chiral superfields $\Phi^N$ belongs to the symmetric product of the adjoint representation of
the underlying gauge group as\\
\begin{eqnarray}
SU(5):    &  (24\times 24)_{symm} = 1+24+75+200 \\ \nonumber
SO(10):   &   (45\times 45)_{symm}=1+54+210+770
\end{eqnarray}

\begin{table}[htb]
\begin{center}
\begin{tabular}{|c|c|}
\hline
\hline
 Representation & $M_{3}:M_{2}:M_{1}$ at $M_{GUT}$ \\
\hline
{\bf 75} of $SU(5)$ & 1:3:(-5) \\
\hline
{\bf 200} of $SU(5)$ & 1:2:10 \\
\hline
{\bf 770} of $SO(10)$: {$H \rightarrow SU(4) \times SU(2) \times SU(2)$} &
1:(2.5):(1.9) \\
\hline
\hline
\end {tabular}
\end{center}
\label{tab:ratio}
\caption {Non-universal gaugino mass ratios for different non-singlet
representations belonging to $SU(5)$ or $SO(10)$ GUT-group that gives rise
to the hierarchy of $M_3 < M_1, M_2$ at the GUT scale.}
\end{table}

Gaugino masses become non-universal if these non-singlet Higgses are responsible for
the GUT-breaking. $75$ and $200$ belonging to
$SU(5)$ or $770$\footnote{For breaking through $770$, we quote
the result, when it breaks through the Pati-Salam gauge group
$G_{422D}$ ($SU(4)_C \times SU(2)_L\times SU(2)_R$ with even
D-parity and assumed to break at the GUT scale itself.} of $SO(10)$
yield a  hierarchy of $M_3 < M_1, M_2$ shown in Table 1. The specific 
non-universal ratio(s) used in the scan can be motivated from GUT breaking
with a linear combination of aforementioned non-singlet representations.

 \noindent {\bf ii) BP2}: Our second benchmark point BP2 is motivated from the Hyperbolic branch/Focus Point 
 region of DM. As has already been mentioned, for mSUGRA, very large values 
 $m_0\sim 10-50$ TeV is required to make $\mu$ small such that LSP becomes 
 predominantly a Higgsino, that paves the way for correct relic abundance through annihilation to 
 $WW$, $ZZ$ and $Ah$ final states. However, introduction of non-universality in the 
 scalar sector, in particular in the Higgs parameters $m_{H_u}$ and $m_{H_d}$ at GUT scale, gives rise to 
 small $\mu$, even without going to such high scalar masses, making it accessible to collider events at LHC.
 Again, following our strategy to minimize the number of parameters to choose BP2, we kept all gaugino and 
 other scalar masses universal at the high scale.

 \noindent {\bf iii) BP3}: Our third benchmark point BP3 represents stau co-annihilation region 
 exploiting non-universality in the scalar sector. We have used squark-slepton non-universality as well as 
 non-universality in the family to make the third family slepton masses lighter than other scalars at the high scale.
 Although such scalar non-universality is mostly phenomenological, 
 having impacts on CP and FCNC issues, it can be motivated from string-inspired models 
 with flavor dependent couplings to the modular fields \cite{3rd1,3rd2}.
 In Table \ref{tab:benchmark} we show all the inputs at high scale as well as the low-scale SUSY masses.

\noindent {\bf iv) MSG}: The mSUGRA benchmark point represent stop co-annihilation region of DM parameter space. 
In mSUGRA, stop co-annihilation occurs at distinct non-zero values of $|A_0|$ in a narrow range, for 
particular values of $m_0,~M_{1/2},~{\rm tan}\beta~{\rm and~Sign}{(\mu)}$.
Higgs mass of 125 GeV can also be obtained in the whole $m_0-M_{1/2}$ plane with 
$m_0> 0.8$ TeV for large $A_0$. Hence, a tiny region of $m_0,~M_{1/2}~{\rm and}~A_0$ 
parameter space simultaneously satisfy right Higgs mass and dark matter constraints. 

However, the situation changes dramatically if we introduce non-universality in gaugino sector, if we assume 
$M_3<M_2=M_1$, effectively adding one more parameter to mSUGRA. Then Higgs mass of 125 GeV can be 
satisfied in a larger range of $A_0$ values; while for a given $A_0$, dark matter density can be satisfied by varying 
$M_{1,2}$ appropriately through stop co-annihilation.

 \begin{figure}[htbp]
\begin{center}
\centerline{\psfig{file=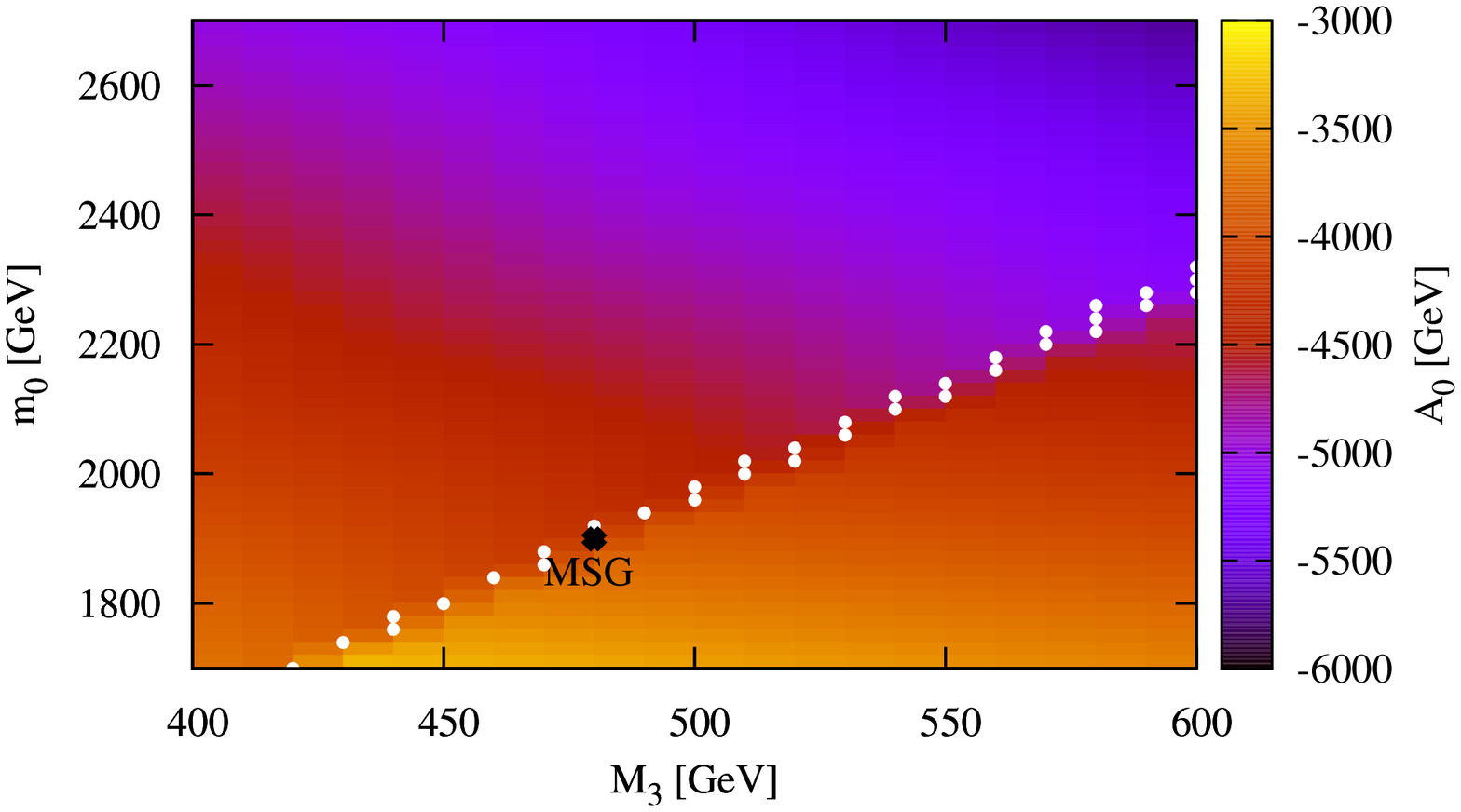,width=6.5 cm,height=7.5cm,angle=-90}
\hskip 20pt \psfig{file=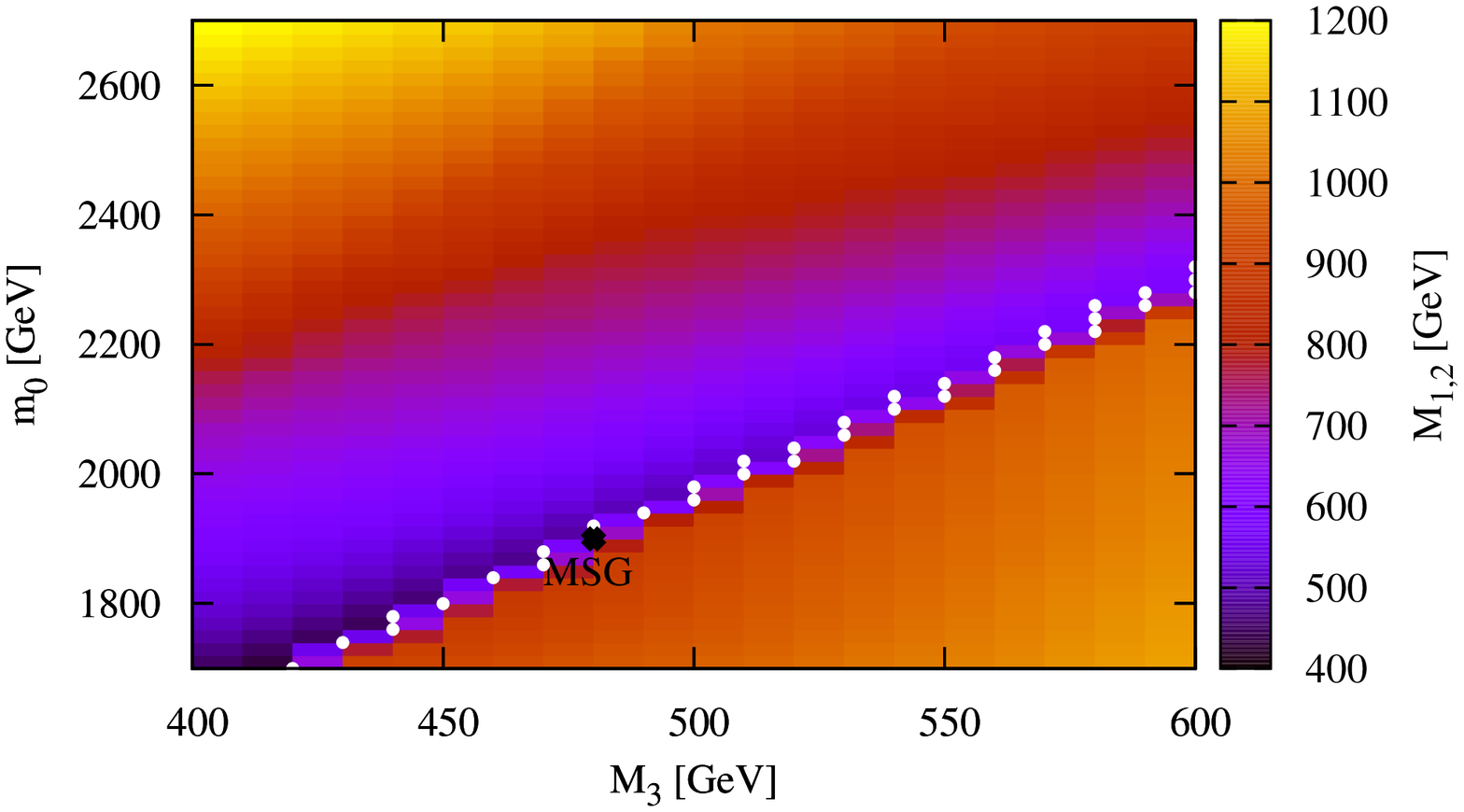,width=6.5 cm,height=7.5cm,angle=-90}}
\caption{Four-dimensional parameter space scan with $m_0,~M_3,~M_{1,2}~{\rm and}~A_0$; 
for ${\rm tan}\beta=15$ and positive $\mu$ to obtain correct dark matter relic abundance, Higgs mass and 
other low energy constraints. LHS: Three-dimensional subset of the scan with $M_3$ (along x-axis), 
$m_0$ (along y-axis), $A_0$ (color gradient); RHS:  $M_3$ (along x-axis), $m_0$ (along y-axis) and
$M_{1,2}$ (color gradient). mSUGRA points are represented in white dots and our benchmark MSG is one of them.} 
\label{fig:MSG}
\end{center}
\end{figure}

In Fig.~\ref{fig:MSG}, 
we have presented a sample scan of such a four-dimensional parameter space $m_0,~M_3,~M_{1,2}~{\rm and}~A_0$, 
for ${\rm tan}\beta=15$ and positive $\mu$. Left panel shows a three-dimensional subset of the scan with $M_3$ (along x-axis), 
$m_0$ (along y-axis), $A_0$ (color gradient) and on the right panel we have $M_3$ (along x-axis), $m_0$ (along y-axis) and
$M_{1,2}$ (color gradient). For a given $M_3$ and $m_0$, there is a range of $A_0$ and $M_{1,2}$ which gives rise to right 
relic abundance and Higgs mass. For simplicity, in Fig.~\ref{fig:MSG}, we consider the minimum possible values of 
$A_0$ and $M_{1,2}$ which are consistent with experimental constraints. As a result, the whole parameter space shown in the 
figure is allowed by dark matter and Higgs mass constraint. White dots in Fig.~\ref{fig:MSG} corresponds to $M_3=M_{1,2}$, 
i.e. mSUGRA points as a subspace of such gaugino non-universality. Our benchmark point MSG is represented by one of these white dots. 
We didn't chose a non-universal benchmark point from this region as the collider signature is expected to be the same as the chosen MSG point.

For renormalization group equation RGE, we use the code {\tt SuSpect v2.3} \cite{suspect} 
with $m_t = 173.2$ GeV, $m_b = 4.2$ GeV, $m_\tau = 1.777$ GeV and stick to 
two-loop RGE with radiative corrections to the gauginos and squarks. We use full one loop and
dominant two loop corrections for the Higgs mass. We ensure radiative electroweak symmetry breaking 
to evaluate the Higgsino parameter $\mu$ at the low scale out of high-scale inputs $m_{H_u}^2$ and $m_{H_d}^2$
and the electroweak symmetry breaking scale has been set
at $\sqrt{m_{\tilde{t_{L}}}m_{\tilde{t_{R}}}}$, the default value in the code
{\tt SuSpect}. The low scale value of the strong
coupling constant has been chosen at
${\alpha_3 (M_{Z})}^{\overline{MS}}= 0.1172$. We compute the cold dark matter relic density with the code micrOMEGAs3.1 \cite {micromegas}.

\begin{table}
\begin{center}
\begin{tabular}{|c|c|c|c|c|}
\hline
\hline
parameter & BP1 & BP2 & BP3 & MSG\\
\hline
\hline
$\tan\beta$ &15.00  &15.00 & 15.00 & 15.00\\
$(M_3,M_2,M_1)$ &(500,700,1282)  &(500,500,500) & (500,500,500) & (480,480,480) \\
$(m_{\widetilde f},m_{\widetilde \tau})$ & (2000,2000) & (2500,2500) & (2000,518) & (1900,1900)\\
$ (m_{H_u},m_{H_d})$ & (2000,2000) & (3047,4000) & (2000,2000) & (1900,1900)\\
$A_0$ & -3700  & -3500 & -3500 & -4239 \\
$sgn(\mu)$ &+  &+ &+ & +\\
\hline
\hline
$m_{\widetilde g}$ & 1251  & 1277 & 1265.2 & 1201.3 \\
$m_{\widetilde u_{L}}$ & 2234  & 2667  & 2217.6 & 2108 \\
$m_{\widetilde t_{1}}$ & 761  & 785.6  & 865 &  243\\
$m_{\widetilde t_{2}}$ & 1656.5  & 1950.2  & 1670 & 1487\\
$m_{\widetilde b_{1}}$ & 1635  & 1940.5  & 1651 & 1442 \\
$m_{\widetilde b_{2}}$ & 2117  & 2558.3  & 2124.6 & 1988 \\
$m_{\widetilde e_{L}}$ & 2054  & 2473  & 2019 &  1918.3\\
$m_{\widetilde \tau_{1}}$ & 1962  & 2420.3  & 219.7 &  1797\\
$m_{\widetilde \tau_{2}}$ & 2013.8 & 2467.2 & 492.2 & 1870 \\
$m_{\widetilde \chi_{1}^{\pm}}$ & 588.3  & 262.6  & 417.6 & 404.6 \\
$m_{\widetilde \chi_{2}^{\pm}}$ & 1584.4  & 447.5  & 1523 & 1742\\
$m_{\widetilde \chi_{4}^{0}}$ & 1584.3  & 447.7  & 1522.4 & 1741 \\
$m_{\widetilde \chi_{3}^{0}}$ & 1581.3  & 285.3  & 1520.3 & 1739.4 \\
$m_{\widetilde \chi_{2}^{0}}$ & 588.4  &  275.3  & 417.6 & 404.6 \\
$m_{\widetilde \chi_{1}^{0}}$ & 561.7  & 201.7  & 211.4 & 208.3\\
\hline
\hline
$m_{h}$ & 124.1 & 123.4 & 123.2 & 123.8\\
$\Omega_{\tilde\chi_1}h^2$& 0.118 & 0.127 & 0.116 & 0.112\\
$BF(b\to s\gamma)$ & $2.98\times 10^{-4}$ & $2.83\times 10^{-4}$
& $3.00\times 10^{-4}$ & $3.25\times 10^{-4}$\\
$Br (B_s \rightarrow \mu^+ \mu^-)$ & $3.10\times 10^{-9}$ & $3.07\times 10^{-9}$
& $3.09\times 10^{-9}$ & $3.13\times 10^{-9}$\\
\hline
\hline
\end{tabular}
\end{center}
\caption {Benchmark points BP1, BP2, BP3 and MSG. Model inputs, low scale
predictions (Masses in GeV) and values of the constraints including Higgs mass and 
relic density are mentioned.}
\label{bechmark-points}
\label{tab:benchmark}
\end{table}

\section {Collider Simulation and Results}

Non-universal SUGRA points advocated in the earlier section 
can be seen at the future run of LHC in bottom rich and 
leptonic final states. This also serves as a major distinguishing feature from mSUGRA points
surviving Higgs mass and dark matter constraints. 

We first discuss the strategy for the simulation including the
final state observables and the cuts employed therein 
and then we discuss the numerical results in next subsection.

\subsection{Strategy for Simulation} 

The spectrum generated by {\tt SuSpect} as described in the earlier
section, at the benchmark points are fed into the event generator
{\tt Pythia} 6.4.16 \cite{Pythia} by {\tt SLHA} interface \cite{sLHA}
for the simulation of $pp$ collision with centre of mass energy 14 TeV for 
LHC.

The default parton distribution functions {\tt CTEQ5L} \cite{CTEQ}, 
QCD scale $\sqrt{\hat{s}}$ in {\tt Pythia} 
has been used. All possible SUSY processes  (mainly 2$\rightarrow$2) and decay chains consistent
with conserved $R$-parity have been kept open with initial and final state radiation on. 
We take hadronization into account using the fragmentation functions
inbuilt in {\tt Pythia}.

The main 'physics objects' that are reconstructed in a collider, are:
\begin{itemize}
\item Isolated leptons identified from electrons and muons
\item Hadronic Jets formed after identifying isolated leptons
\item Unclustered Energy made of calorimeter clusters with $p_T~>$ 0.5 GeV
(ATLAS) and $|\eta|<5$, not associated to any of the above types of
high-$E_T$ objects (jets or isolated leptons).
\end{itemize}
We try to mimic the experimental reconstruction for these objects 
in Pythia as follows.
\bei
\item {\em Isolated leptons} ($\ell$):
\eei
Isolated leptons are identified as electrons and muons with $p_T>$ 10 GeV
and $|\eta|<$2.5. An isolated lepton is separated from another lepton by 
${\bigtriangleup R}_{\ell\ell}~ \geq $0.2, from jet 
(jets with $E_T >$ 20 GeV) with ${\bigtriangleup R}_{{\ell}j}~ \geq 0.4$, while 
the energy deposit $\sum {E_{T}}$ due to low-$E_T$ hadron activity around a
lepton within $\bigtriangleup R~ \leq 0.2$ of the lepton axis should be
$\leq$ 10 GeV. $\bigtriangleup R = \sqrt {{\bigtriangleup \eta}^2
+ {\bigtriangleup \phi}^2}$ is the separation in pseudo rapidity and
azimuthal angle plane. The smearing functions of isolated electrons, photons
and muons are described below.

\bei
\item {\em Jets} ($jet$):
\eei Jets are formed with all the final
state particles after removing the isolated leptons from the list
with {\tt PYCELL}, an inbuilt cluster routine in {\tt Pythia}. The
detector is assumed to stretch within the pseudorapidity range
$|\eta|$ from -5 to +5 and is segmented in 100 pseudorapidity
($\eta$) bins and 64 azimuthal ($\phi$) bins. The minimum $E_T$ of
each cell is considered as 0.5 GeV, while the minimum $E_T$ for a
cell to act as a jet initiator is taken as 2 GeV. All the partons
within $\bigtriangleup R$=0.4 from the jet initiator cell is
considered for the jet formation and the minimum $\sum_{parton}
{E_{T}}^{jet}$ for a collected cell to be considered as a jet is
taken to be 20 GeV. We have used the smearing function and
parameters for jets that are used in {\tt PYCELL} in {\tt Pythia}.

\bei
\item {\em b-jets}:
\eei
We identify partonic $b$ jets by simple $b$-tagging algorithm with
efficiency of $\epsilon_b = 0.5$ for $p_T >$ 40 GeV and $|\eta| <$ 2.5 
\cite{b-tagging-ref}.

\bei
\item {\em Unclustered Objects} ($Unc.O$):
\eei
 All the other final state
particles, which are not isolated leptons and separated from jets by
$\bigtriangleup R \ge$0.4 are considered as unclustered objects.
This clearly means all the particles (electron/photon/muon) with
$0.5< E_T< 10$GeV and $|\eta|< 5$ (for muon-like track $|\eta|<
2.5$) and jets with $0.5< E_T< 20$GeV and $|\eta|< 5$, which are
detected at the detector, are considered as unclustered objects.

\begin{itemize}

\item {Electron/Photon Energy Resolution :}

\be
\sigma(E)/E=a/\sqrt{E}\oplus b\oplus c/E  \footnote{$\oplus$ indicates addition in quadrature}
\ee

Where,\\
$~~~~~~~~$ $a$ = 0.03 [GeV$^{1/2}$], $~~$   $b$ = 0.005 \& $~$   $c$ = 0.2 [GeV] 
$~~~$  for $|\eta|< 1.5$ \\     
$~~~~~~~~~~~~$= 0.055 $~~~~~~~~~~~~~~~~$ = 0.005  $~~~~~~~$ = 0.6 $~~~~~~~~~$  
for $1.5< |\eta|< 5$ \\

\item { Muon $P_T$ Resolution :}

\begin{eqnarray}
\sigma(P_T)/P_T&=&a  ~~~~~~~~~~~~~~~~~~~~~~~{\rm if} ~ P_T< 100 GeV\\
&=&a+b\log(P_T/\xi) ~~~~~ {\rm if}~ P_T> 100 GeV
\end{eqnarray}

Where,\\
$~~~~~~~~$ $a$= 0.008 $~~$\&    $b$= 0.037 $~~~~~$  for $|\eta|< 1.5$ \\     
$~~~~~~~~~~$= 0.02 $~~~~~~$    = 0.05  $~~~~~~~~~$ $1.5< |\eta|<2.5$\\
and $\xi=100$ GeV.

\item {Jet Energy Resolution :}

\be
\sigma(E_T)/E_T=a/\sqrt{E_T}
\ee

Where,\\
$~~~~~~~~$ a= 0.55 [GeV$^{1/2}$], default value used in {\tt PYCELL}.

\item {Unclustered Energy Resolution :}

\be
\sigma(E_T)=\a\sqrt{\Sigma_{i}E^{(Unc.O)i}_T}
\ee

Where, $\a\approx0.55$. One should keep in mind that the x and y component of 
$E^{Unc. O}_T$ need to be smeared independently with same smearing 
parameter.

\end{itemize}

We sum vectorially the x and y components of the momenta separately for all visible
objects to form visible transverse momentum $(p_T)_{vis}$,
\bea
(p_T)_{vis}=\sqrt{(\sum p_x)^2+(\sum p_y)^2}
\eea
where, $\sum p_x =\sum (p_x)_{iso~\ell}+\sum (p_x)_{jet}+\sum (p_x)_{Unc.O}$
and similarly for $\sum p_y$.
We identify $(p_T)_{vis}$ as missing energy
$E_{T}\!\!\!\!/$:
\bea
E_{T}\!\!\!\!/ = (p_T)_{vis}
\eea

We also define Effective mass $H_T$ as the scalar sum of transverse momenta of 
visible objects like lepton and jets with missing energy 

\bea
H_T=\sum {p_T}^{\ell_i}+ {p_T}^{jets}+{E_{T}}\!\!\!\!/ 
\eea

Effective mass cuts have really been useful to reduce SM background for 
the signals as we will see shortly.

We studied the benchmark points in multi-lepton final states as well as in 
b-rich final states at $E_{CM}$= 14 TeV at LHC with varying cuts. The 
channels we study are:

\begin{itemize}

\item Four b-jet with inclusive lepton and jets $(4b)$ :
$4b~+ X~ + {E_{T}}\!\!\!\!/$ ; Here $X$ implies any number of inclusive jets or leptons without any specific veto on that. 
 Basic cuts applied here are ${p_T}^b>$ 40 GeV,  ${E_{T}}\!\!\!\!/ >$100 GeV.

\item Four b-jet with single lepton $(4b\ell)$ :
 $4b~+ \ell~ + X~ + {E_{T}}\!\!\!\!/$ ; Here $X$ implies any number of inclusive jets without any specific veto on that. The lepton can have any charge $\pm$.
 Basic cuts applied here are ${p_T}^b>$ 40 GeV, ${p_T}^{\ell}>$ 20 GeV, $|\eta| <$ 2.5, ${E_{T}}\!\!\!\!/ >$100 GeV.

\item Two b-jets with di-lepton $(2b2\ell)$:
$2b~ + 2\ell~ + X~ + {E_{T}}\!\!\!\!/$ ; Here $X$ implies any number of inclusive jets without any specific veto on that. Leptons can have any charge $\pm$ 
(including same and opposite sign). Basic cuts applied here are ${p_T}^b>$ 40 GeV, ${p_T}^{\ell}>$ 20 GeV, $|\eta| <$ 2.5, ${E_{T}}\!\!\!\!/ >$100 GeV.

\item Same sign dilepton with inclusive jets $(\ell^{\pm}\ell^{\pm})$: 
$\ell^{\pm}\ell^{\pm}~+X~ + {E_{T}}\!\!\!\!/$ ;
The basic cuts applied are ${E_{T}}\!\!\!\!/ >$ 30 GeV, ${p_T}^{\ell_1}>$ 40 GeV and ${p_T}^{\ell_2} >$ 30 GeV with $|\eta| <$ 2.5.

\item Trilepton with inclusive jets ($\ell^{\pm}\ell^{\pm}\ell^{\pm}$): 
$\ell^{\pm}\ell^{\pm}\ell^{\pm}~+X~ + {E_{T}}\!\!\!\!/$ ;
Basic cuts ${E_{T}}\!\!\!\!/ >$ 30 GeV, ${p_T}^{\ell_1} >$ 30 GeV, ${p_T}^{\ell_2} >$ 30 GeV and ${p_T}^{\ell_3} >$ 20 GeV with $|\eta| <$ 2.5.

\item Four-lepton with inclusive jets ($\ell^{\pm}\ell^{\pm}\ell^{\pm}$): 
$\ell^{\pm}\ell^{\pm}\ell^{\pm}\ell^{\pm}~+X~ + {E_{T}}\!\!\!\!/$ ;
For basic cuts no missing energy cut is employed while, lepton transverse momentum cuts are as follows: ${p_T}^{\ell_1}>$ 20 GeV, ${p_T}^{\ell_2} >$ 20 GeV and ${p_T}^{\ell_3} >$ 20 GeV and ${p_T}^{\ell_4} >$ 20 GeV with $|\eta|<$ 2.5.
\end{itemize}

$\ell$ stands for final state isolated electrons and or muons as discussed above and 
${E_{T}}\!\!\!\!/$ depicts the missing energy. Opposite-sign dilepton was not considered mainly because of the huge SM background from $t\bar{t}$ process. 

Apart from the basic cuts including a Z-veto of $|M_Z-M_{\ell^+\ell^-}|\ge$15 GeV on same flavor opposite sign dilepton arising in $2b2l$, trilepton and 
four lepton final states, we apply sum of lepton $p_T$ cut ($\sum {p_T}^{\ell_i}$) and combination of lepton  $p_T$ cut with MET, called modified effective mass cut 
$H_{T1}=\sum {p_T}^{\ell_i}+{E_{T}}\!\!\!\!/ $ to the leptonic final states, and harder $H_T$ cuts on b-rich final states and we refer to them as follows: 

\begin{itemize}
\item $C1$: $\sum {p_T}^{\ell_i}>$ 200 GeV
\item $C2$: $\sum {p_T}^{\ell_i}>$ 400 GeV
\item $C3$: $H_{T1} >$ 400 GeV
\item $C4$: $H_{T1} >$ 500 GeV
\item $C1'$: $\sum {p_T}^{\ell_i}>$ 100 GeV
\item $C2'$: $\sum {p_T}^{\ell_i}>$ 200 GeV
\item $C3'$: $H_{T1} >$ 150 GeV
\item $C4'$: $H_{T1} >$ 250 GeV
\item$C5$: $H_T >$ 1000 GeV, ${E_{T}}\!\!\!\!/ >$ 200 GeV, ${p_T}^b>$ 60 GeV.
\end{itemize}

 We have generated dominant SM events from $t\bar t$ in {\tt Pythia} for the same final 
states with same cuts and multiplied the corresponding events in different channels 
by proper $K$-factor ($1.59$) to obtain the usually noted next to leading order 
(NLO) and next to leading log re summed (NLL) cross-section at LHC \cite{ttbar}. 
$b\bar{b}b\bar{b}$ ,$b\bar{b}b\bar{b}W/Z$ and $t\bar{t}b\bar{b}$ background have 
been calculated in {\tt Madgraph5}\cite{Madgraph}.
The cuts are motivated such that we reduce the background to a great extent as shown 
in next subsection. Note that softer cuts $C1',C2',C3',C4'$ have been used 
for four lepton channel where the SM background is much smaller. 

\subsection{Numerical results}
\begin{table}[!ht]
\begin{center}
\begin{tabular}{|c|r|r|r|r|r|r|r|r|r|r}
\hline
Model Points & \multicolumn{1}{c|}{Total} &
  \multicolumn{1}{c|}{$\tilde g \tilde g$} &
\multicolumn{1}{c|}{$\tilde t_1 {\tilde t_1}^*$} 
& \multicolumn{1}{c|}{$\tilde{\chi}_i^0 \tilde{\chi}_j^0$} 
& \multicolumn{1}{c|}{$\tilde{\chi}_i^{\pm} \tilde{\chi}_j^{\mp}$} 
& \multicolumn{1}{c|}{$\tilde{\chi}_i^{0} \tilde{\chi}_j^{\pm}$} 
& \multicolumn{1}{c|}{$\tilde g \rightarrow \tilde t_1 \bar{t} $}
& \multicolumn{1}{c|}{$\tilde t_1 \rightarrow t \chi_1^0$} & 
\multicolumn{1}{c|}{$\tilde t_1 \rightarrow b \chi_1^{+}$}\\
\hline
\hline
 {\bf BP1} & 107.6 & 29.20 & 32.06& 0.11 & 7.18 & 14.6 & 99 $\%$ & 76.3 $\%$ & 23.7 $\%$\\
\hline
 {\bf BP2} & 607 & 26.3 & 15.1 & 64.7 & 126.9 & 354.8 & 99 $\%$ &10.3$\%$ & 45.2$\%$\\
\hline
 {\bf BP3} & 188 & 26.5 & 13.8 & 0.33 & 37.1 & 74.2 & 99 $\%$& 86.5$\%$ & 9.4$\%$  \\
\hline
 {\bf MSG} &18208 & 39  &  18010  &0.1 &28 &84 & 99$\%$ & 0$\%$ & 0$\%$\\
\hline
\hline
\end {tabular}
\end{center}
\caption {Total supersymmetric particle production cross-sections
(in fb) as well as some leading contributions from  $\tilde g \tilde g$ and $\tilde t_1 {\tilde t_1}^*$ and electroweak 
neutralino-chargino productions for each of the benchmark points with $E_{CM}$= 14 TeV. 
We also quote the significant decay branching fractions (in percentage). }
\label{production}
\end{table}

The main SUSY production cross-sections for the benchmark points have been noted in Table \ref{production} with the total 
cross-section for all 2$\rightarrow$2 SUSY processes.  All the non-universal benchmark points have 
similar gluino production and third family stop production, while the mSUGRA point has a huge stop production 
due to very light stop mass and the total cross-section for this point is also dominated by that.
Although other benchmark points have sufficiently large branching fraction of stop going to 
bottom chargino or stop neutralino, MSG has nothing in these channels as the stop is almost degenerate 
with the lightest neutralino, it only decays to $c \tilde{\chi}_1^{0}$ in loop.  For MSG, $\tilde{\chi}_2^{0}$ decays to 
$\tilde{\chi}_1^{0}h$ 95$\%$ and first chargino dominantly decays to $\tilde{t}_1\bar{b}$. Hence 3b channel can be a better 
channel to look for such MSG points. As mSUGRA is only alive in such a region of parameter space for the sake of dark matter, 
all MSG points will be similar in this aspect. We also note that for  BP1: $\tilde{\chi}_1^{\pm}$ decays into $\ell + \nu_{\ell}+ \tilde{\chi}_1^{0}$ 
through off-shell sleptons in $33\%$ while $\tilde{\chi}_2^{0}$ decays to leptonic final state is only $\simeq 1\%$.
BP2 has dominant production in electroweak gauginos. Associated production of the gluinos with neutralinos are also quite heavy. 
Here $\tilde t_1 \rightarrow t \tilde{\chi}_{2,3}^0$ branchings are also of the same order of $\tilde t_1 \rightarrow t \tilde{\chi}_{1}^0$. 
Although $\tilde{\chi}_2^{0}$ decays to  leptonic final state is 1$\%$, 
$\tilde{\chi}_1^{\pm}$ decays into $\ell + \nu_{\ell}+ \tilde{\chi}_1^{0}$ in 33$\%$. Huge electroweak production will significantly 
contribute to leptonic final states for BP2. For BP3, chargino and neutralino decays to tau-rich final state as a 
result of lighter stau. Hence, in addition to the standard leptons, channels with tau-tagging can be a better channel to look for 
this benchmark point.

\begin{figure}[htbp]
\begin{center}
\centerline{\psfig{file=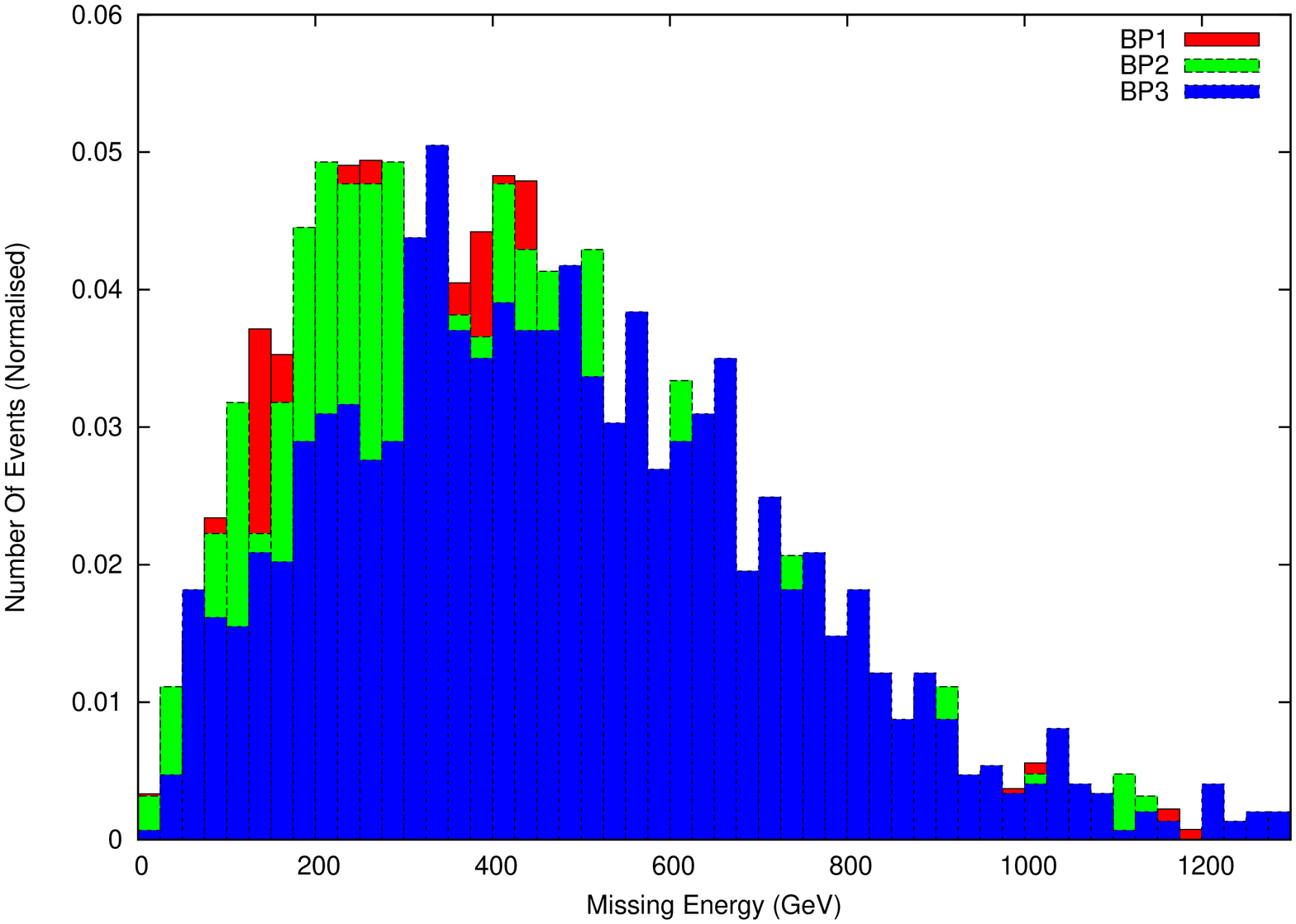,width=6.5 cm,height=7.5cm,angle=-90}
\hskip 20pt \psfig{file=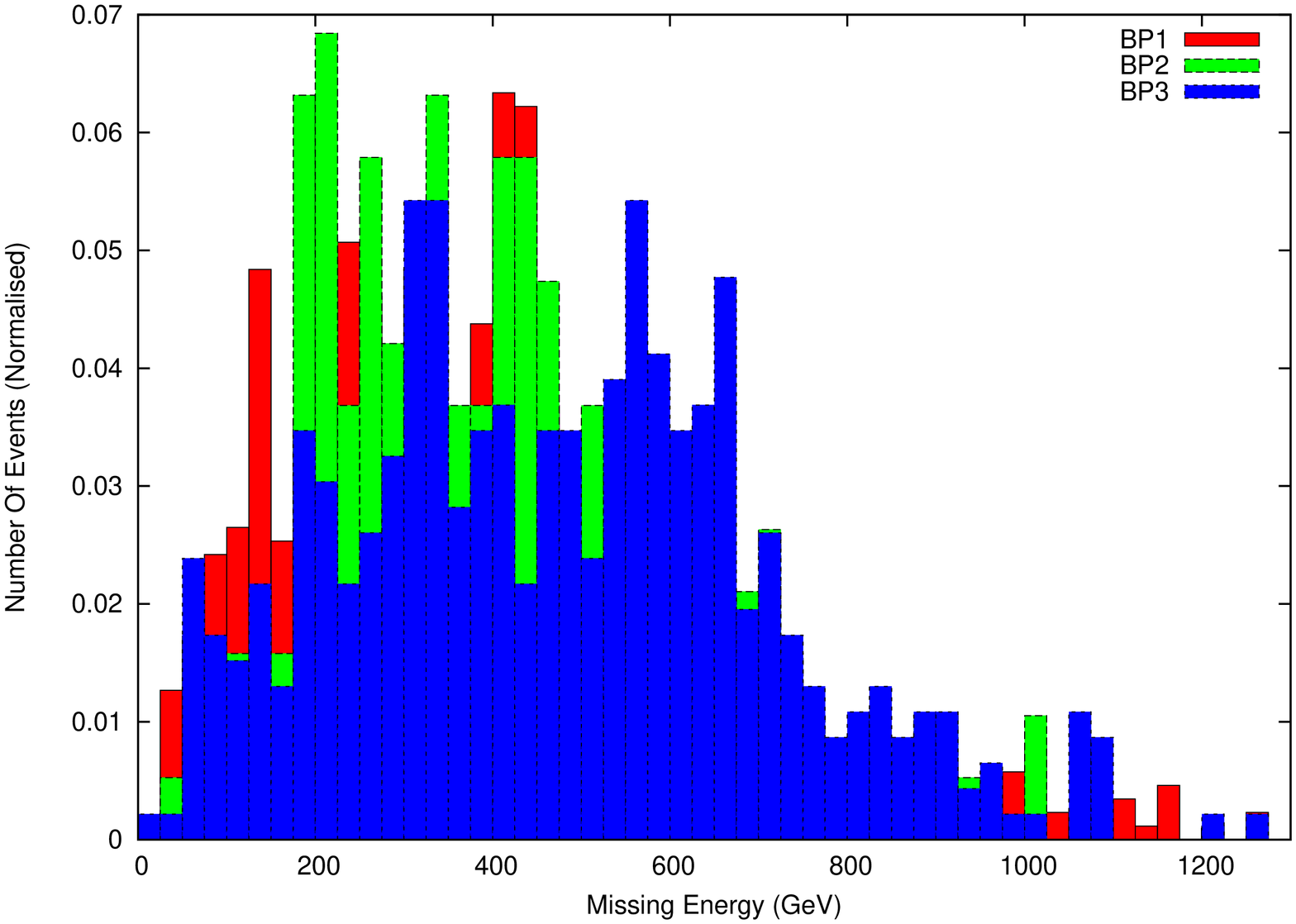,width=6.5 cm,height=7.5cm,angle=-90}}
\vskip 10pt
\centerline{\psfig{file=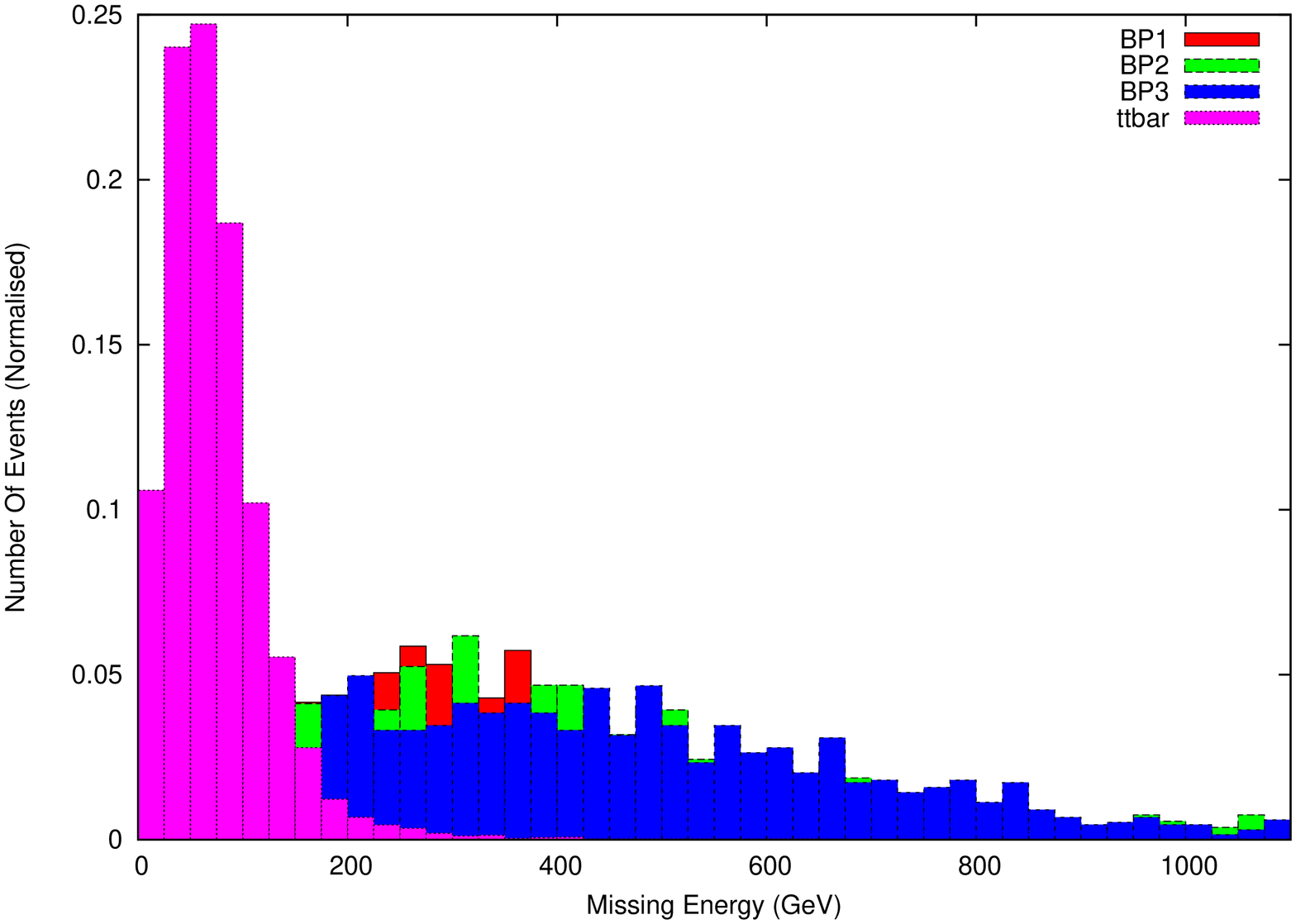,width=6.5 cm,height=7.5cm,angle=-90}}
\caption{Missing energy distribution in bottom rich final states at the benchmark points.
Top left: $4b$ channel, Top right: $4b\ell$ channel; bottom: $2b2\ell$ channel.
{\tt CTEQ5L} pdfset was used. Factorization and 
Renormalization scale has been set to $\mu_F=\mu_R=\sqrt{\hat s}$, 
sub-process centre of mass energy.}
\label{fig:MET1}
\end{center}
\end{figure}

\begin{figure}[htbp]
\begin{center}
\centerline{\psfig{file=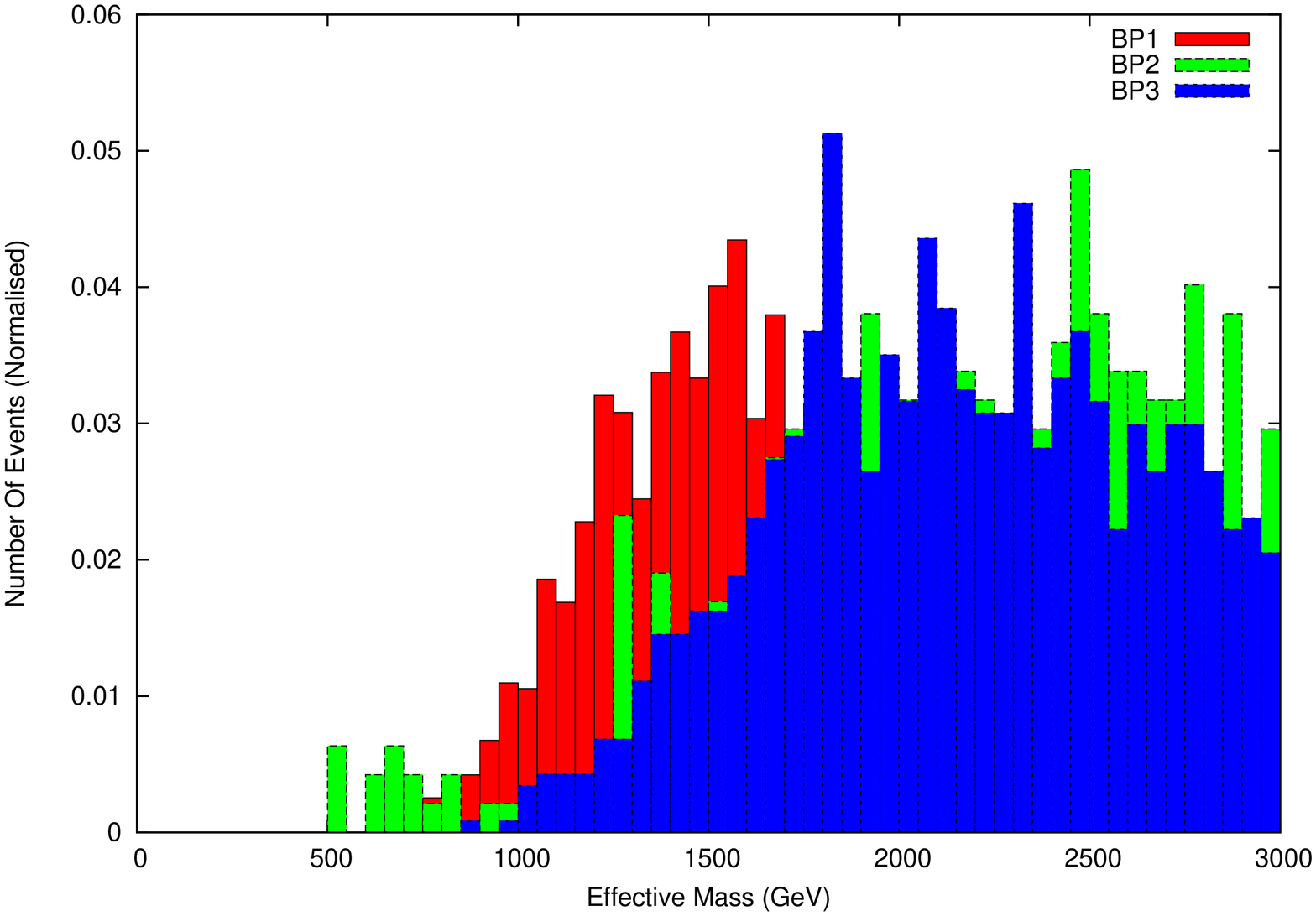,width=6.5 cm,height=7.5cm,angle=-90}
\hskip 20pt \psfig{file=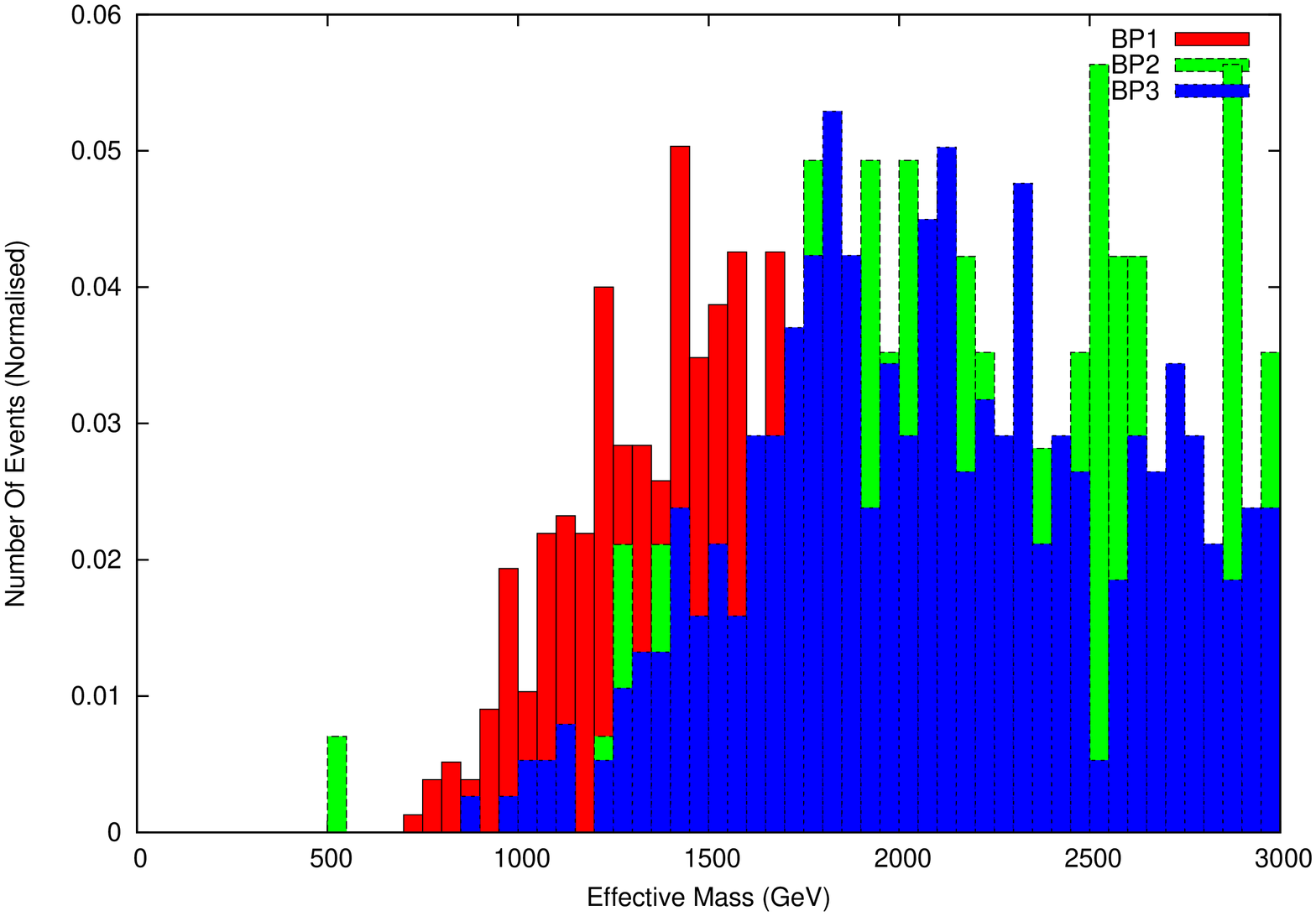,width=6.5 cm,height=7.5cm,angle=-90}}
\vskip 10pt
\centerline{\psfig{file=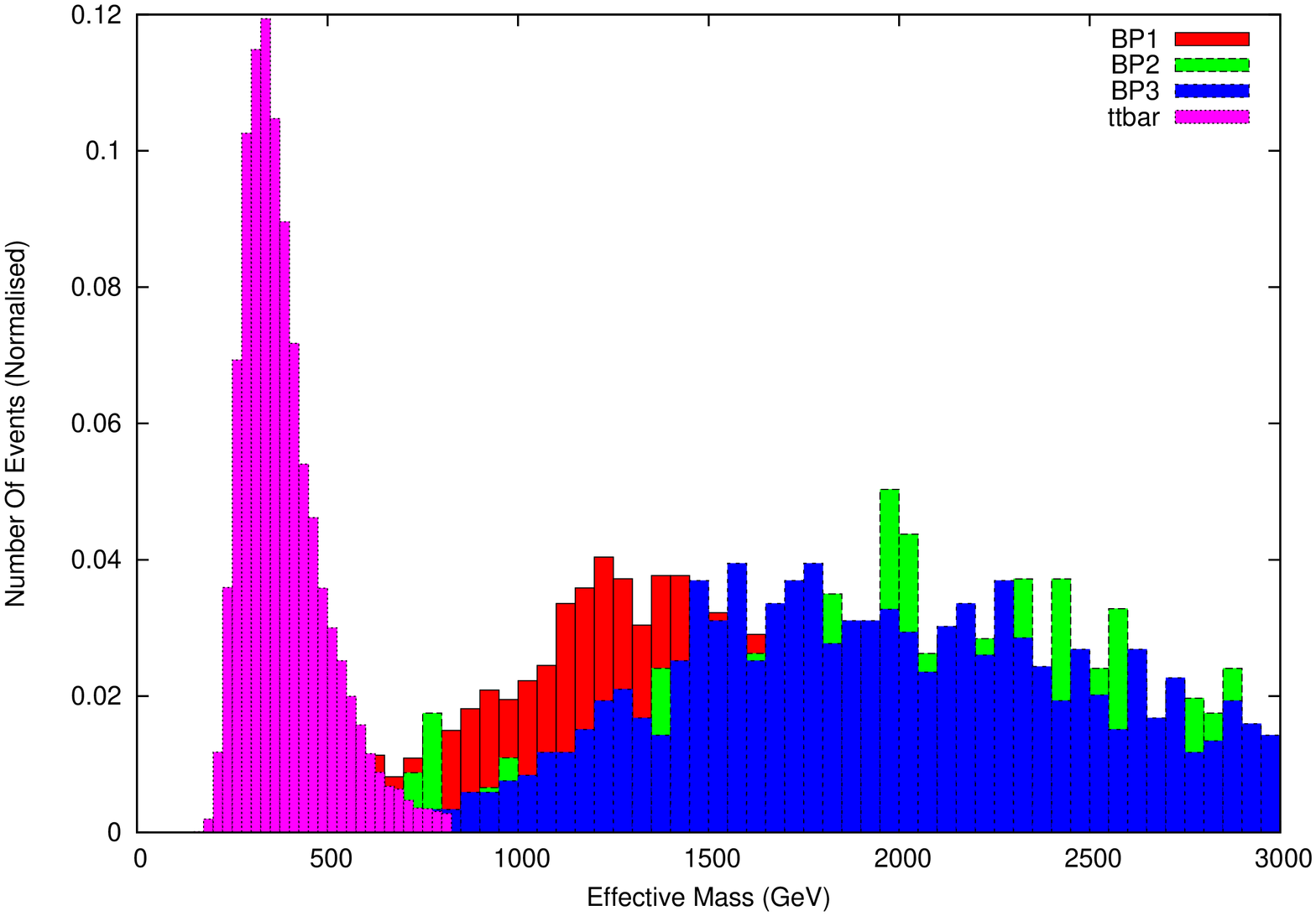,width=6.5 cm,height=7.5cm,angle=-90}}
\caption{ Effective mass distribution in bottom rich final states at the benchmark points. 
Top left: $4b$ channel, Top right: $4b\ell$ channel; bottom: $2b2\ell$ channel.
{\tt CTEQ5L} pdfset was used. Factorization and 
Renormalization scale has been set to $\mu_F=\mu_R=\sqrt{\hat s}$, 
sub-process centre of mass energy.}
\label{fig:EFT1}
\end{center}
\end{figure} 

\begin{table}
\begin{center}
\begin{tabular}{|c|c|c|c|c|c|c|c|c|}
\hline
Benchmark Points & $\sigma_{4b}$ & $\sigma_{4bl}$ &$\sigma_{2b2l}$ &$\sigma_{4b}$($C5$) & $\sigma_{4bl}$($C5$) &$\sigma_{2b2l}$($C5$)\\
\hline
\hline
BP1 & 1.35 & 0.44 & 1.15 & 0.60 & 0.18 & 0.84\\
\hline
BP2 & 1.56 & 0.50 & 1.24 & 1.53 & 0.49 & 1.11\\
\hline
BP3 & 1.34 & 0.41 & 1.17 & 0.76 & 0.22 & 0.91 \\
\hline
MSG & 0.004 & 0.004  &  0.1 & $\le$0.001 & $\le$0.001  &  0.01 \\
\hline
\hline
$t\bar{t}$ & $\le$0.01 & $\le$0.01 & 973.1 & $\le$0.01 & $\le$0.01 & $\le$0.01 \\
\hline
$b\bar{b}b\bar{b},b\bar{b}b\bar{b} + W/Z $& 0.106 & $\le$0.01 & $\le$0.01 & $\le$0.01 & $\le$0.01 & $\le$0.01\\
\hline
$t\bar{t}b\bar{b}$ &0.8825 & 0.634 & 1.03 & 0.005 & $\le$0.01  & $\le$0.01 \\
\hline
\hline
\end {tabular}
\end{center}
\vspace{0.2cm}
\caption{ Event-rates (fb) in bottom rich final states at the chosen
benchmark points for $E_{CM}$= 14 TeV with basic cuts and cuts $C5$. {\tt CTEQ5L} pdfset was
used. Factorization and Renormalization scale has been set to 
$\mu_F=\mu_R=\sqrt{\hat s}$, subprocess centre of mass energy. Contributions from dominant SM 
backgrounds are also noted. } 
\label{b-events}
\end{table}

Missing energy distribution of the benchmark points in bottom rich final states 
are shown in figure \ref{fig:MET1}. Missing Energy has been normalized to 1. $4b$ and $4b\ell$ 
final states doesn't have a significant background, hence only signal events are shown. It occurs 
that the benchmark points have a similar missing energy pattern, while for $2b2\ell$, the $t\bar{t}$ background has a 
sharper peak at low missing energy as can be expected. Similarly effective mass  $H_T$ distribution in bottom-rich final states 
is shown in figure \ref{fig:EFT1}. 
There is no significant difference between the benchmark points in terms of this distribution either. 
We can see for $4b\ell$ channel (Fig \ref{fig:MET1}, top right), the peaks of the distributions are a bit separated. 
For $2b2\ell$, background $t\bar{t}$ peaks at a much lower value while the signal events have a peak $\ge$ 1000 GeV. 
This gives us the opportunity to put a very hard effective mass $H_T$ cut, which reduces the background to almost zero, 
while retaining the signal. Hard effective mass cut also helps to remove other hadronic and QCD backgrounds as 
shown in Table \ref{b-events}.   

In summary, from Table \ref{b-events}, BP1, BP2 and BP3 have very good prospects of being 
discovered at LHC in $4b$, $4b\ell$ and $2b2\ell$ final states while the corresponding MSG 
point doesn't contribute at all in such final states. The main reason of this is clear from 
Table \ref{production}. Although $\tilde t_1 {\tilde t_1}^*$ 
production is huge for MSG, stop being almost degenerate with LSP, 
it can not decay to $t \tilde{\chi_1^0}$ or $b \tilde{\chi_1^{+}}$ and hence
it doesn't produce any $b$-jets. We might however, see 3$b$ events from 
electroweak production. 

\begin{figure}[htbp]
\begin{center}
\centerline{\psfig{file=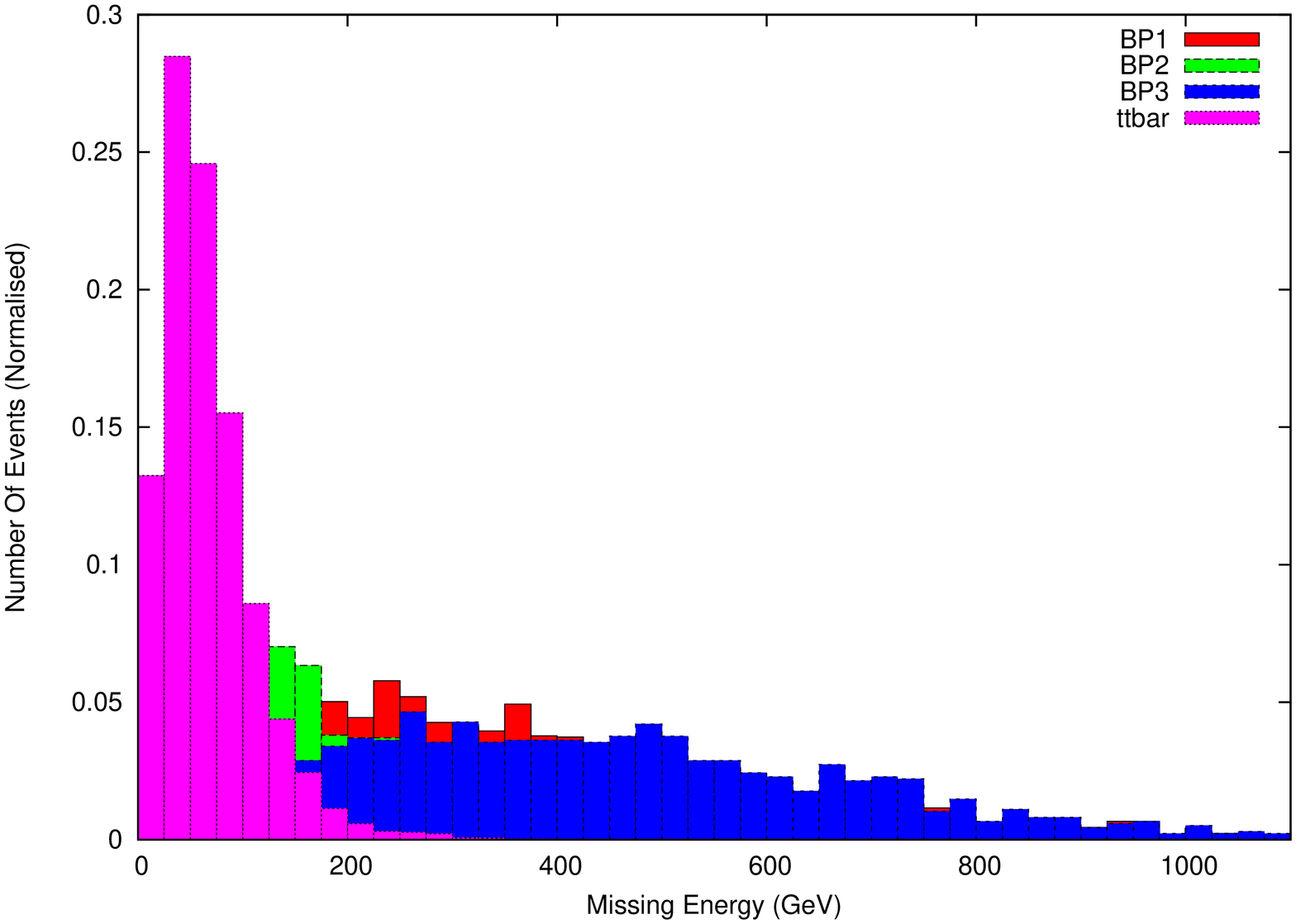,width=6.5 cm,height=7.5cm,angle=-90}
\hskip 20pt \psfig{file=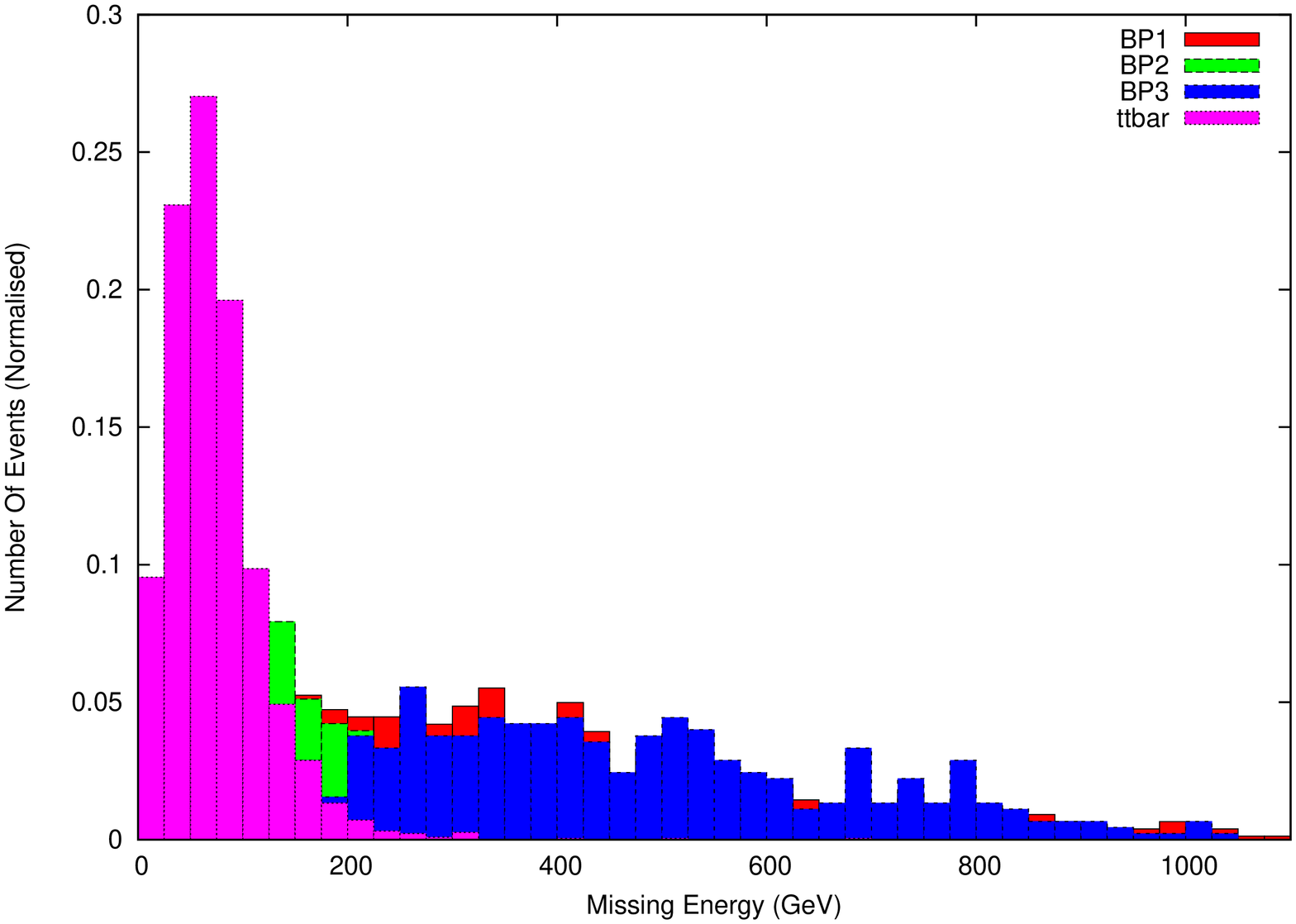,width=6.5 cm,height=7.5cm,angle=-90}}
\caption{Missing energy distribution in  $\ell^{\pm}\ell^{\pm}$ (left) and 
$\ell^{\pm}\ell^{\pm}\ell^{\pm}$ (right) final states at the benchmark points.}
\label{fig:MET2}
\end{center}
\end{figure} 

\begin{figure}[htbp]
\begin{center}
\centerline{\psfig{file=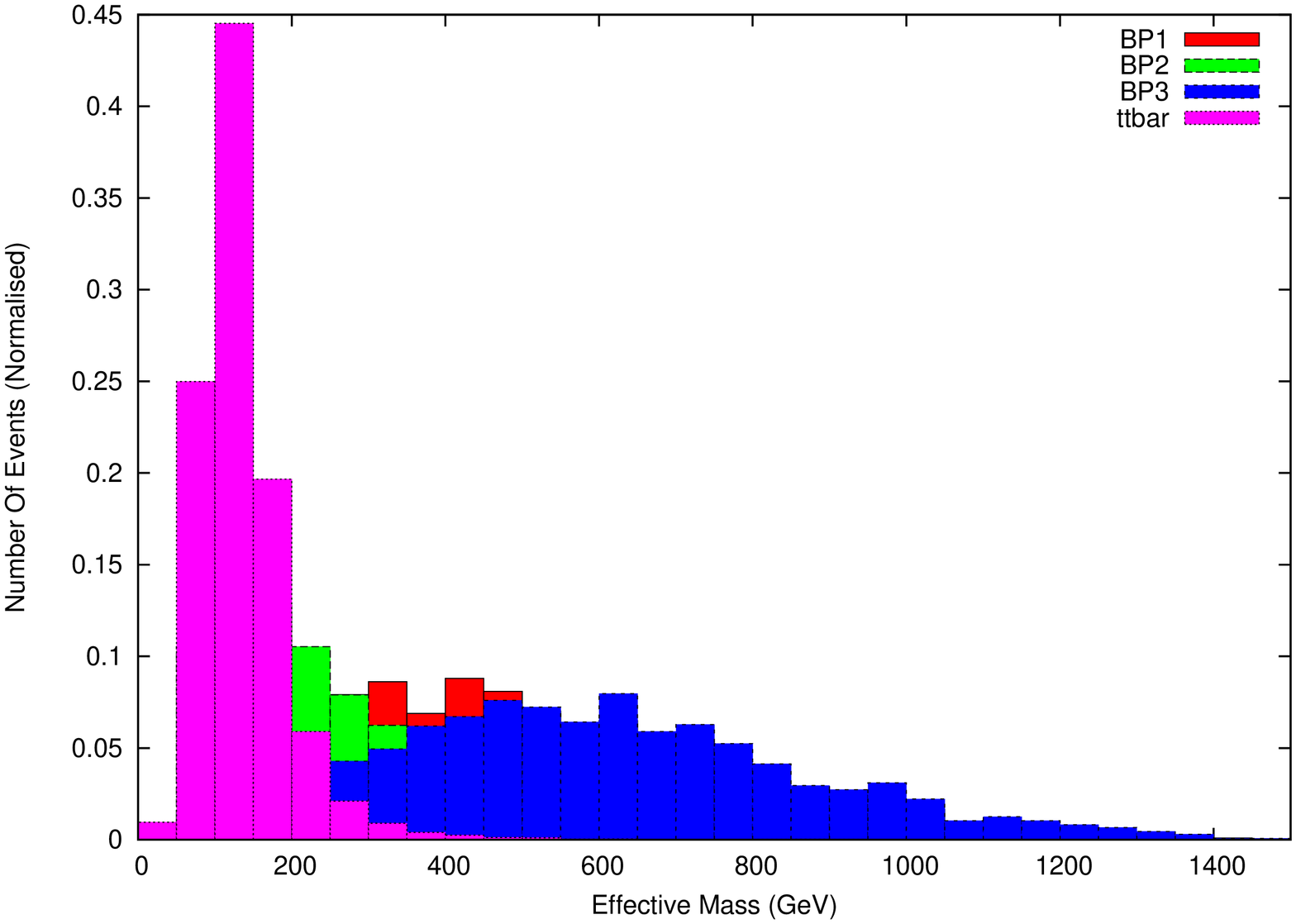,width=6.5 cm,height=7.5cm,angle=-90}
\hskip 20pt \psfig{file=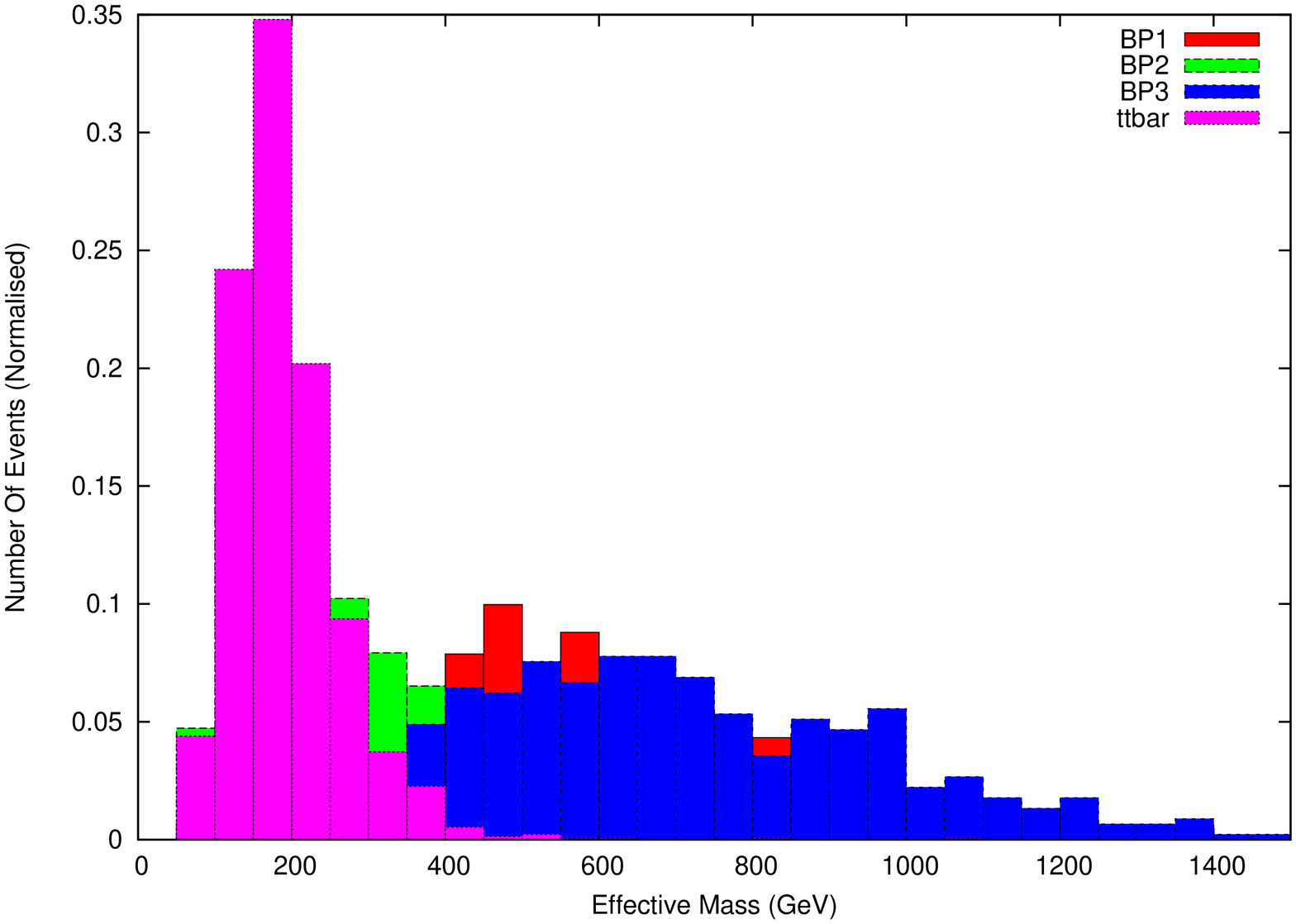,width=6.5 cm,height=7.5cm,angle=-90}}
\caption{Effective mass $H_{T_1}$ distribution in  $\ell^{\pm}\ell^{\pm}$ (left) and 
$\ell^{\pm}\ell^{\pm}\ell^{\pm}$ (right) final states at the benchmark points.} 
\label{fig:EFT2}
\end{center}
\end{figure}

The SM backgrounds are negligible in bottom rich channels excepting $2b2\ell$, which suffers from  
a sufficiently large background from $t\bar{t}$ production. But, a heavy Effective mass cut ($H_T$) eliminates this to 
a large extent, while retaining the signals. The Effective mass distribution in Fig. \ref{fig:EFT1} bears the testimony 
to the fact. We also note that SM background events were simulated with very high 
number of events, such that each event carries a small weight, 0.01 fb of cross-section; hence, null 
events in simulation corresponds to cross-section less than that.

\begin{table}
\begin{center}
\begin{tabular}[ht]{|c|c|c|c|c|c|}
\hline
\hline
Channels and Event rates ($fb$) & BP1 & BP2 & BP3 & MSG & $t\bar{t}$\\
\hline
\hline
 $\ell^{\pm}\ell^{\pm}$ (Basic) & 0.48  & 1.03 & 0.65 &0.2 & 40.32\\
 $\ell^{\pm}\ell^{\pm}$+$C1$& 0.16  & 0.30 & 0.30 & 0.1 & 1.08\\
 $\ell^{\pm}\ell^{\pm}$+$C2$& 0.03 & 0.03 & 0.05& $\le$0.001 & $\le$0.01\\
 $\ell^{\pm}\ell^{\pm}$+$C3$& 0.35 & 0.53 & 0.54 & 0.1 & 0.54\\
$\ell^{\pm}\ell^{\pm}$+$C4$& 0.26 & 0.40 &  0.44 & $\le$0.001 & 0.17\\
\hline
$\ell^{\pm}\ell^{\pm}\ell^{\pm}$ (Basic) & 0.18 &  0.96 & 0.24 & $\le$0.001 & 33.96\\
 $\ell^{\pm}\ell^{\pm}\ell^{\pm}$+$C1$& 0.11  & 0.40 & 0.18 & $\le$0.001 & 3.62\\
$\ell^{\pm}\ell^{\pm}\ell^{\pm}$+$C2$& 0.02 & 0.06 & 0.04 & $\le$0.001 & 0.17\\
$\ell^{\pm}\ell^{\pm}\ell^{\pm}$+$C3$& 0.15 & 0.38 & 0.22 & $\le$0.001 & 0.54\\
$\ell^{\pm}\ell^{\pm}\ell^{\pm}$+$C4$& 0.11 & 0.31 &  0.19 & $\le$0.001 & $\le$0.01\\
\hline
$\ell^{\pm}\ell^{\pm}\ell^{\pm}\ell^{\pm}$ (Basic) & 0.018  & 0.21 & 0.019& $\le$0.001 & 0.17\\
$\ell^{\pm}\ell^{\pm}\ell^{\pm}\ell^{\pm}$+$C1'$& 0.018  & 0.20 & 0.019& $\le$0.001 & 0.17\\
$\ell^{\pm}\ell^{\pm}\ell^{\pm}\ell^{\pm}$+$C2'$& 0.013 & 0.11 & 0.017& $\le$0.001 & $\le$0.01\\
$\ell^{\pm}\ell^{\pm}\ell^{\pm}\ell^{\pm}$+$C3'$& 0.017 & 0.21 & 0.019 & $\le$0.001 & 0.17\\
$\ell^{\pm}\ell^{\pm}\ell^{\pm}\ell^{\pm}$+$C4'$& 0.016 & 0.16 &  0.019 & $\le$0.001 & $\le$0.01\\
\hline
\end{tabular}
\end{center}
\caption {Event-rates (fb) in leptonic final states at the chosen
benchmark points for $E_{CM}$= 14 TeV with basic cuts and cuts $C1$, $C2$, $C3$, $C4$ as described. 
The main background $t\bar{t}$ is also noted. {\tt CTEQ5L} pdfset was used. Factorization and Renormalization scale has been set to
$\mu_F=\mu_R=\sqrt{\hat s}$, subprocess centre of mass energy. Note that trilepton and four-lepton final states include $Z-$veto.}
\label{lep-events}
\end{table}

Missing energy and effective mass distribution for Same-sign dilepton and trilepton 
events are shown in fig \ref{fig:MET2} and \ref{fig:EFT2} respectively. Again all the benchmark 
points show very similar distribution, while the $t \bar{t}$ can be reduced with a heavy $H_{T_1}$ cut.
All the leptonic event numbers for the benchmark points are shown in Table \ref{lep-events}. 

Table \ref{lep-events} tells us, that trilepton events are 
still good for all the benchmark points while 4-lepton channel is good for BP2 and BP3 only. 
We also need to note that the background for 4-lepton channel is negligible 
(hadronically quiet part comes from $4W$ or $ZZZ$ production). After the cuts they vanish almost completely. 
Similarly $ZW$, which contributes to trilepton reduces to a great extent after the Z-veto. 
Hence, we didn't quote those background events here. We also see that $C2$ and $C4$ cut
reduce the $t\bar{t}$ background significantly. 
$C2$ kills the signal events to a great extent too, hence, $C4$ is a better choice to 
reduce background and retain signal. Hence, these leptonic final states are also good channels to study 
such benchmark points. The reason of BP2 having larger leptonic events, 
comes also from huge electroweak gaugino productions as pointed in Table \ref{production}. 
Hence, a significant part of these leptonic final states should contain hadronically quiet lepton events. 
The minimal supergravity benchmark point doesn't contribute at all to the leptonic final states, the reason 
being simply understood as not having lighter stops to decay through top or sleptons leading  to leptons. 
Hence, such mSUGRA points can only be studied in hadronic channels or perhaps $3b$ final states as mentioned earlier. 
After mSUGRA being alive only in stop co-annihilation region, this seems to be a generic feature 
for all mSUGRA parameter space points to obey Higgs mass and dark matter constraint.  
This in turn, can help distinguishing such non-universal frameworks from mSUGRA in LHC signature space.

\section {Summary and Conclusions}

It is remarkable that a Higgs boson has been discovered with a mass $\simeq 125$ GeV. In pure SM, theoretically
there is no reason why its mass should be at the EW scale, or even it is, why it is not much higher or lower
than 125 GeV. (In fact, in pure SM,  best fit to the EW data prefers a much lower mass). This gives us hope
that some symmetry principle is there beyond the pure SM, and supersymmetry being the most natural candidate,
because it solves the hierarchy problem, as well as it constraints the Higgs mass to be less than $\sim 135$ GeV.
In addition, supersymmetry has a natural candidate for the dark matter. However, the minimal version of the 
most desirable version of MSSM, mSUGRA, is in very tight corner to satisfy all the existing experimental
constraints, as well as being within the reach of LHC. We find that mSUGRA is still viable in the stop co-annihilation region
in which the classic SUSY signal (multijet plus missing $E_T$) is essentially unobservable beyond the SM background
at the LHC. (The other allowed region such as hyperbolic/ focus point has SUSY particle masses well beyond the
reach of LHC). However, if we relax little bit from mSUGRA with non-universal gaugino and /or scalar masses, 
the situation becomes much more favorable to discover SUSY at the LHC.

In this work, we have shown that SUSY with non-universalities in gaugino or scalar  masses within high scale SUGRA
set up can still be accessible at LHC with $E_{CM}=$ 14 TeV. In particular, 
we show the consistency of the parameter space in different dark matter annihilation regions. 
Wino dominated LSP with chargino co-annihilation can be achieved with gaugino mass 
non-universality with $M_3<M_2<M_1$. Hyperbolic Branch/Focus point 
region with Higgsino dominated LSP can be obtained easily with Higgs non-universality as BP2. 
Such parameter space automatically occurs with lighter gauginos and hence they may dominate 
the production and leptonic final states at LHC. Stau co annihilation can occur with scalar non-universality 
while stop co annihilation can arise simply with high-scale gaugino non-universality with $M_3<M_2=M_1$. 
mSUGRA, though viable in only stop co-annihilation region, do not yield lepton or b-rich final states due to 
lack of phase space for the stop to decay leptonically. There exist a reasonable region of parameter space
 in the non-universal scenario which not only satisfy all the existing constraints, but also can unravel SUSY in bottom and 
lepton rich final states with third family squarks being lighter than the first two automatically. We have made detailed studied
of three benchmark points in these allowed parameter spaces, and find that SUSY signal in the bottom or bottom plus 
lepton-rich final state  stands over the SM  background with suitable cuts. We have also investigated pure leptonic final states
with suitable cuts, and find some of these final state have viable prospects. 
Finally we also emphasize that  with good luminosity in the upcoming 14 TeV LHC runs, these allowed parameter space can be 
ruled out easily, or we we will discover SUSY.

{\bf Acknowledgment:} The work of SB is supported by the U.S. Department of Energy under 
Grant No. DE-SC0008541. The work of SC, KG and SN was supported in part by the U.S. Department of Energy
Grant Number DE-SC0010108.





\end{document}